\documentclass[preprint,nofootinbib,aps,superscriptaddress,eqsecnum]{revtex4-2} 
\usepackage[colorlinks,citecolor=purple]{hyperref}
\usepackage[usenames]{color}
\usepackage{xcolor}
\usepackage{float}
\pdfoutput=1

\usepackage{tabularx}
\usepackage{graphicx}
\usepackage{amsmath}
\usepackage{amsfonts}
\usepackage{amssymb}
\usepackage[justification=centering]{caption} 
\usepackage{caption}
\usepackage{subcaption}
\usepackage{color}
\allowdisplaybreaks
\def\bea{\begin{eqnarray}}
	\def\eea{\end{eqnarray}}
\def\be{\begin{equation}}
	\def\ee{\end{equation}}

\begin{document}
	\title{Multi-component Dark Matter and leptogenesis with double seesaw in an extended left-right symmetric theory}
	\author{Ankita Kakoti}
	\email{kakotiankita97@gmail.com}
	\affiliation{Department of Physics, Sibsagar University, Sibsagar, 785665, India}
	\begin{abstract}
		Left-Right Symmetric theory has proved to be one of the most successful models in explaining the origin of neutrino mass and mixings, and the phenomenological origin of neutrinoless double beta decay, lepton flavor violation as well as leptogenesis within its regime. In the current work, left-right symmetric model has been extended with a sterile neutrino per generation, which acts as a dark matter candidate within the model. The model has been realized using $A_{4}$ modular symmetry and the primary focus of the work rests on investigating the impact of assigning different modular weights to the particle content of the model and hence study its effects on the results pertaining to relic abundance of dark matter and also leptogenesis within its framework.
	\end{abstract}

	\maketitle
	\newpage
	
	\section{Introduction}
	\label{w71}
	
	With tremendous advancement in research in the area of particle physics and cosmology, various suggestions regarding the presence of a non-luminuous, non-baryonic form of matter in the Universe has gained the forefront. This matter is often referred to as Dark Matter(DM) \cite{Zwicky:1933gu}. According to Planck data, about $26.8\%$ of the energy density of the Universe comprises of DM. The abundance of DM is stated in terms of the density parameter $\Omega_{DM}$ and reduced Hubble constant, $h$, which is generally given as \cite{Planck:2018vyg},
	\begin{equation}
		\label{7e1}
		\Omega_{DM}h^{2}=0.012 \pm 0.002
	\end{equation}
	However, whether dark matter is actually existent in the Universe is still a question as the Standard Model (SM) of particle physics does not put forward any suitable candidate for DM.\\
	Another most intriguing puzzle in the observation of matter dominated Universe, which points towards the matter-antimatter asymmetry present in our Universe. This is often referred to as baryon asymmetry of the Universe(BAU). For baryon asymmetry to take place, the three conditions put forward by Andrei Sakharov must be satisfied, which are namely 1) Baryon number violation, 2) C and CP violation and 3) Out of equilibrium decays \cite{Sakharov:1988vdp}. Although a small amount of CP violation is also found in the SM, but the CP violation in the quark sector is considerably minimal and falls short by almost ten orders of magnitude. So, BAU cannot be successfully explained by the most successful SM of particle physics. The observed BAU is generally stated in terms of baryon to photon ratio, which is given by,
	\begin{equation}
		\label{7e2}
		\frac{n_{B}-n_{\bar{B}}}{n_{\gamma}} \approx 10^{-10}
	\end{equation}\\
	Finding answers to these questions demanded particle physicists to think of something beyond the SM framework. One such framework is the left-right symmetric model (LRSM) \cite{Grimus:1993fx,BhupalDev:2018xya,Lee:2017mfg}, which is an extension of SM with a $SU(2)_{R}$ gauge group and incorporation of $U(1)_{B-L}$. LRSM in this work has however been extended by a sterile neutrino per generation, which will act as a suitable dark matter candidate within the model. Scalar sector of the model consists of a Higgs bidoublet $\phi(1,2,2,0)$ and two doublets $H_{L}(1,2,1,-1)$ and $H_{R}(1,1,2,-1)$.  As a result of this incorporation, the resulting light neutrino mass in LRSM which otherwise is the summation of type-I and type-II seesaw masses will now be given by the double seesaw mechanism \cite{Grimus:2013tva}.\\
	The model has been realized using $A_{4}$ modular symmetry, where the primary focus of the work is to analyse the impact of assigning varying modular weights to the particle content of the model. Considering $A_{4}$ modular symmetry, which is isomorphic to $\Gamma_{3}$ modular group, the Yukawa couplings will be expressed in terms of modular forms of a particular weight. In one of our previous work, we have studied keV sterile neutrino dark matter within modular LRSM with $k_{Y} = 2$. However, under the $A_{4}$ group, assigning different weight to the modular forms can result in different irreducible representations of the said modular form which in turn can result in different neutrino mass structures in the context of the model.\\
	In section \ref{w72} of the paper, an introduction to doublet LRSM with double seesaw mechanism has been put forward, in section \ref{w73}, a brief discussion on sterile neutrino dark matter and leptogenesis has been taken into consideration. In section \ref{w74}, $\Gamma_{3}$ modular symmetry and irreducible representations corresponding to different modular weights has been discussed. Section \ref{w75} puts forward the explanation regarding the use of different modular weights corresponding to $\Gamma_{3}$ modular group and the resulting neutrino mass matrices. In section \ref{w76}, the numerical analysis and results of the present work has been discussed and finally section \ref{w77} gives the thorough discussion of the results obtained and hence concluding the work. 
	\section{Doublet LRSM and double seesaw mechanism}\label{w72}
	The scalar sector of LRSM can consist of the following fields,\\
	\begin{itemize}
		\item one Higgs bidoublet $\phi(1,2,2,0)$ and two scalar triplets $\Delta_L(1,3,1,2)$ and $\Delta_R(1,1,3,2)$ \cite{Maiezza:2016ybz,Senjanovic:2016bya}.
		\item one Higgs bidoublet $\phi(1,2,2,0)$ and two scalar doublets $H_{L}(1,2,1,-1)$ and $H_{R}(1,1,2,-1)$ \cite{Senjanovic:1978ev,Mohapatra:1977be,Borah:2025fkd}.
	\end{itemize}
	In context of the present work, where the light neutrino mass originates by the double seesaw mechanism, the scalar sector of the model consists of a Higgs bidoublet and two scalar doublets, that is, the work focusses on doublet left-right symmetric model\cite{Das:2016akd,Das:2017hmg}.\\
	In some works it has been suggested that the scalar triplet $\Delta_{R}$ can act as a suitable DM candidate, but this work extends LRSM with a sterile neutrino per generation which acts as the DM candidate. The light neutrino mass within the model will be determined by the Double Seesaw mechanism \cite{Patel:2023voj}. Here, the fermion sector will consist of the usual quarks and leptons and the scalar sector will consist of one Higgs bidoublet and two scalar doublets, as shown below,\\
	\textbf{Fermion sector}
	\begin{equation}
		\label{7e3}
		q_{L} = \begin{pmatrix}
			u_{L}\\
			d_{L}
		\end{pmatrix} \sim (3,2,1,1/3),
		q_{R} = \begin{pmatrix}
			u_{R}\\
			d_{R}
		\end{pmatrix} \sim (3,1,2,1/3)
	\end{equation}
	\begin{equation}
		\label{7e4}
		l_{L} = \begin{pmatrix}
			\nu_{L}\\
			e_{L}
		\end{pmatrix} \sim (1,2,1,-1),
		l_{R} = \begin{pmatrix}
			\nu_{R}\\
			e_{R}
		\end{pmatrix} \sim (1,1,2,-1)
	\end{equation}
	\textbf{Higgs sector}
	\begin{equation}
		\label{7e5}
		H_{L} = \begin{pmatrix}
			h_{L}^{+}\\
			h_{L}^{0}
		\end{pmatrix} \sim (1,2,1,-1),
		H_{R} = \begin{pmatrix}
			h_{R}^{+}\\
			h_{R}^{0}
		\end{pmatrix} \sim (1,1,2,-1),
		\phi = \begin{pmatrix}
			\phi_{1}^{0} & \phi_{2}^{+}\\
			\phi_{1}^{+} & \phi_{2}^{0} 
		\end{pmatrix}\sim (1,2,2,0)
	\end{equation}
	
	The superpotential associated with the model is given by,
	\begin{equation}
		\label{7e6}
		\mathcal{W} = l_{L}^{T}i\sigma_{2}\phi l_{R}^{C}Y_{LR}+fY_{RS}l_{R}^{T}i\sigma_{2}H_{R}S_{L}+\mu S_{L}S_{L}
	\end{equation}
	\begin{equation}
		\label{7e7}
		\supset M_{D}\nu_{L}^{T}N_{R}+M_{RS} N_{R}S_{L}+\mu S_{L}S_{L}
	\end{equation}
	The spontaneous symmetry breaking (SSB) in the model takes place in two steps, namely,
	\begin{center}
		\centering
		$ SU(3)_{C} \otimes SU(2)_{L} \otimes SU(2)_{R} \otimes U(1)_{B-L}$\\
		$\Bigg\downarrow$ \hspace{0.5cm} $v_{R}$\\
		$ SU(3)_{C} \otimes SU(2)_{L}  \otimes U(1)_{Y}$\\
		$\Bigg\downarrow$ \hspace{0.3cm} $v$\\
		$SU(3)_{C} \otimes U(1)_{em}$
	\end{center}
	where, $v$ is the vev of the Higgs bidoublet $\phi$ and $v_{R}$ is the vev of the scalar doublets $H_{R}$ . It is to be noted here that the scalar doublet $H_{L}$ does not take part in spontaneous symmetry breaking and its presence only caters to the left-right invariance.  As a consequence, the left-handed scalar doublet does not get any vev, that is, $<H_{L}>=0$ \cite{Kakoti:2025eub}.\\
	After SSB, the neutrino mass matrix is depicted as,
	\begin{equation}
		\label{7e8}
		\mathcal{M} =\begin{pmatrix}
			0 & M_{D} & 0 \\
			M_{D}^{T} & 0 & M_{RS}\\
			0 & M_{RS}^{T} & \mu
		\end{pmatrix}
	\end{equation}
	Following the mass hierarchy $M_{D}$, $M_{RS}$ $>>$ $\mu$, 
	\begin{equation}
		\label{7e9}
		m_{\nu} = M_{D}(M_{RS}^{T})^{-1}\mu M_{RS}^{-1}M_{D}^{T}
	\end{equation}
	where, $M_{D}$ is the Dirac mass matrix, $M_{RS}$ is the active-sterile mixing matrix and $\mu$ is the sterile neutrino Majorana mass matrix. As is evident from the expression \eqref{7e9}, the resulting light neutrino mass is doubly suppressed by the active-sterile neutrino mixing matrix. The double seesaw mechanism is performed by incorporating the type-I seesaw mechanism twice.  So, for the fact that $M_{R}$ $>>$ $M_{D}$, we get the mass of the heavy neutrinos in the form \cite{Patra:2023ltl},
	\begin{equation}
		\label{7e10}
		M_{R} = M_{RS}.\mu^{-1}M_{RS}^{T}
	\end{equation}
	which is the implication of the first seesaw. The second seesaw shows that the resulting neutrino mass matrix becomes linearly dependent on the sterile neutrino mass $\mu$ as shown in expression \eqref{7e9}. However, as is also evident from the said equation that the light neutrino mass is doubly suppressed by the active-sterile mixing matrix and hence the named has been dubbed as, 'double seesaw mechanism' \cite{Grimus:2013tva}.
	\section {Dark Matter and Leptogenesis in doublet LRSM}\label{w73}
	\subsection{Sterile Neutrino Dark Matter}\label{ww73}
	In this work, LRSM has been extended with a sterile neutrino per generation. In many cases, the lightest of the same is considered to be the dark matter candidate \cite{Dodelson:1993je,Boyarsky:2018tvu}. What's different in this case is the fact that the three sterile neutrinos are degenerate in mass which points towards a multi-component dark matter. The active-sterile mixing has also been found to be almost degenerate as depicted in the results described below. \\
	One of the most important criteria for being a DM candidate is that it has to be stable, at least on the cosmological scale. However, the sterile neutrinos incorporated into the model are not perfectly stable, they can decay radiatively to a photon and an active neutrino via the process $S \rightarrow \nu + \gamma$, which produces an X-ray signal. However, as stated in many literature \cite{Drewes:2016upu}, the decay rate is negligible owing to the small mixing between active and sterile states. Now, since the sterile neutrino is a fermionic dark matter candidate, lower bound exists on its mass known as the Tremaine-Gunn bound \cite{Boubekeur:2023fqo,Davoudiasl:2020uig}. An upper limit on the dark matter mass is also obtained from the X-ray constraints and as a matter of fact, direct and indirect detection of DM also imposes significant constraints on the sterile neutrino as a DM candidate.\\
	Any stable neutrino state having non-zero mixing with the active neutrino state is produced through the active-sterile neutrino conversion. Thus, the DM abundance is produced through the mixing of the active-sterile neutrino states. The mechanism of non-resonant production of sterile neutrino is known as the Dodelson-Widrow(DW) mechanism \cite{Merle:2015vzu}.\\
	To calculate the relic abundance as well as decay rate of the dark matter candidate, we have used formulae stated in equations \eqref{7e11} and \eqref{7e12}.
	The relic abundance corresponding to a dark matter candidate can be given by the expression \cite{Asaka:2006nq},\\
	\begin{equation}
		\label{7e11}
		\Omega_{DM}h^{2} = 1.1 \times 10^{7} \Sigma C_{\alpha} (m_{s})|U_{\alpha s}|^{2}\Bigg(\frac{m_{s}}{keV}\Bigg)^{2}, \alpha = e , \mu , \tau
	\end{equation}
	where $C_{\alpha}(m_s)$ represents active flavor dependent coefficients, which can numerically be computed by solving Boltzmann equations.
	Equation \eqref{7e11} can be simplified to,
	\begin{equation}
		\label{7e12}
		\Omega_{DM}h^{2} \approxeq 0.3\Bigg(\frac{sin^{2} 2\theta_{DM}}{10^{-10}}\Bigg)\Bigg(\frac{m_{s}}{100keV}\Bigg)^{2}
	\end{equation}
	where, $sin^{2}\theta_{DM} = 4\sum_{\alpha = e,\mu,\tau}|U_{\alpha s}|^{2}$, $|U_{\alpha s}|$ being the active-sterile neutrino mixing and $m_{s}$ is the mass of the lightest sterile neutrino which is the DM candidate.\\
	However, as all the three sterile neutrinos will be contributing towards the phenomenology of dark matter, the resulting dark matter relic abundance will be given as,
	\begin{equation}
		\label{7e13}
		\Omega_{DM}h^{2}= \Omega_{S_{1}}h^{2}+\Omega_{S_{2}}h^{2}+\Omega_{S_{3}}h^{2}
	\end{equation}
	The decay rate is given by \cite{Gautam:2019pce},
	\begin{equation}
		\label{7e14}
		\Gamma = 1.38 \times 10^{-32}\Bigg(\frac{sin^{2} 2\theta_{DM}}{10^{-10}}\Bigg)\Bigg(\frac{m_{s}}{100keV}\Bigg)^{5} s^{-1}
	\end{equation}
	Again, the total decay rate will be expressed as,
	\begin{equation}
		\label{7e15}
		\Gamma_{total}=\Gamma_{1}+\Gamma_{2}+\Gamma_{3}
	\end{equation}
	From  equations \eqref{7e12} and \eqref{7e14} it is evident that the relic abundance and decay rate both depend upon the mixing of active-sterile states and mass of the DM candidate. Hence, the same set of model parameters that can produce correct neutrino phenomenology can also be used to evaluate the relic abundance and the decay rate of the sterile neutrino.
	\subsection{Leptogenesis in doublet LRSM}\label{www73}
 Considering the presence of three right-handed neutrinos in the present scenario, an unflavored analysis (considering on decay of $N_{1}$) for CP-asymmetry has been defined as \cite{Davidson:2008bu,Buchmuller:2005eh,DiBari:2012fz,Nir:2007zq},
 \begin{equation}
 	\epsilon_{1} = \frac{\Gamma(N_{1}\rightarrow l\phi)-\Gamma(N_{1}\rightarrow l^{c}\phi^{c})}{\Gamma_{D}}
 \end{equation}
 where, $\Gamma_{D}$ is the total decay rate for $N_{1}$ and it can be expressed as,
 \begin{equation}
 	\label{g1}
 	\Gamma_{D}=\Gamma(N_{1}\rightarrow l\phi)+\Gamma(N_{1}\rightarrow l^{c}\phi^{c})
 	                       =\frac{[Y_{D}^{\dagger}Y_{D}]_{11}m_{N_{1}}}{8\pi}
 \end{equation}
 It is known that the interaction rate which are slower than the Hubble expansion rate are enough for bringing particle distributions to equilibrium, which in result leas to the following condition,
	\begin{equation}
		\label{7e16}
		\Gamma_{D} \leq H (T=M_{N_{1}})
	\end{equation}
	where, $\Gamma_{D}$ is given by the equation \eqref{g1}.
and, $Y_{D}=\frac{M_{D}}{v}$ is the Yukawa coupling.
Working under the assumption that, the RH neutrinos follow the mass hierarchy, $M_{N_{1}}<M_{N_{2}}<M_{N_{3}}$, such that the decay of $N_{1}$ essentially determines the sought-after asymmetry. \\
	\section{$\Gamma_{3}$ modular group and different modular weights}\label{w74}
	The concept of modular symmetry \cite{Feruglio:2017spp,King:2020qaj} has put forward a way to study generation of neutrino masses within a model without the use of extra particles or flavons. In this work we have realized LRSM with $A_{4}$ modular symmetry where the model contains usual particle content, without the use of any flavons \cite{Sahu:2020tqe}. While using $A_{4}$ modular symmetry, the Yukawa couplings are expressed in terms of modular forms with each form being assigned a particular modular weight. Different weights of the modular forms under $A_{4}$ can have different irreducible representations and as such, each weight will result in a different neutrino mass texture and hence variations in the results for phenomenology under consideration. Different irreducible representations for different weights under $A_{4}$ modular symmetry has been discussed as under \cite{Ding:2022aoe,Lu:2019vgm,Zhang:2019ngf}.
	\begin{itemize}
		\item Considering the weight of the modular group as 2, that is, $k_{Y}=2$, will give us a triplet representation of $Y$ that is, there will be three number of modular forms given as,
		\begin{equation}
			\label{7e21}
			Y_{(3)}^{2} = \begin{pmatrix}
				Y_1 \\
				Y_2 \\
				Y_3
			\end{pmatrix}
		\end{equation}
		where, the subscript within brackets represents the multiplet and the superscript stands for the corresponding weight under $A_4$.\\
		\item For $k_{Y}=4$, there will be five number of modular forms, two singlets $1$, $1'$ and one triplet $3$ under $A_{4}$, represented as,
		\begin{equation}
			\label{7e22}
			Y_{(1)}^{4} = Y_{1}^{2} + 2 Y_{2} Y_{3}
		\end{equation}
		\begin{equation}
			\label{7e23}
			Y_{(1')}^{4} = Y_{3}^{2} + 2 Y_{1} Y_{2}
		\end{equation}
		\begin{equation}
			\label{7e24}
			Y_{(3)}^{4} = \begin{pmatrix}
				Y_{1}^{2}-Y_{2}Y_{3}\\
				Y_{3}^{2}-Y_{1}Y_{2}\\
				Y_{2}^{2}-Y_{1}Y_{3}
			\end{pmatrix}
		\end{equation}
		\item For $k_{Y}=6$, there will be seven number of modular forms, two triplets $3_{1}$, $3_{2}$ and one singlet $1$ under $A_{4}$, represented as,
		\begin{equation}
			\label{7e25}
			Y_{(1)}^{6} = Y_{1}^{3}+Y_{2}^{3}+Y_{3}^{3}-3Y_{1}Y_{2}Y_{3}
		\end{equation}
		\begin{equation}
			\label{7e26}
			Y_{(3_{1})}^{6} = Y_{1}^{2}+2Y_{2}Y_{3}\begin{pmatrix}
				Y_{1}\\
				Y_{2}\\
				Y_{3}
			\end{pmatrix}
		\end{equation}
		\begin{equation}
			\label{7e27}
			Y_{(3_{2})}^{6} = Y_{3}^{2}+2Y_{1}Y_{2}\begin{pmatrix}
				Y_{3}\\
				Y_{1}\\
				Y_{2}
			\end{pmatrix}
		\end{equation}
		\item For $k_{Y}=8$, there will be nine number of modular forms, three singlets $1$,$1'$,$1''$ and two triplets $3_{1}$, $3_{2}$ under $A_{4}$, represented as,
		\begin{equation}
			\label{7e28}
			Y_{(1)}^{8} = (Y_{1}^{2}+2Y_{2}Y_{3})^{2}
		\end{equation}
		\begin{equation}
			\label{7e29}
			Y_{(1')}^{8} = (Y_{1}^{2}+2Y_{2}Y_{3})(Y_{3}^{2}+2Y_{1}Y_{2})
		\end{equation}
		\begin{equation}
			\label{7e30}
			Y_{(1'')}^{8} = (Y_{3}^{2}+2Y_{1}Y_{3})^{2}
		\end{equation}
		\begin{equation}
			\label{7e31}
			Y_{(3_{1})}^{8} = (Y_{1}^{2}+2Y_{2}Y_{3})\begin{pmatrix}
				Y_{1}^{2}-Y_{2}Y_{3}\\
				Y_{3}^{2}-Y_{1}Y_{2}\\
				Y_{2}^{2}-Y_{1}Y_{3}
			\end{pmatrix}
		\end{equation}
		\begin{equation}
			\label{7e32}
			Y_{(3_{2})}^{8} = (Y_{3}^{2}+2Y_{1}Y_{2})\begin{pmatrix}
				Y_{2}^{2}-Y_{1}Y_{3}\\
				Y_{1}^{2}-Y_{2}Y_{3}\\
				Y_{3}^{2}-Y_{1}Y_{2}
			\end{pmatrix}
		\end{equation}
		\item For $k_{Y}=10$, there will be nine number of modular forms, three singlets $1$,$1'$ and three triplets $3_{1}$, $3_{2}$ and $3_{3}$ under $A_{4}$, represented as,
		\begin{equation}
			\label{7e33}
			Y_{(1)}^{10} = (Y_{1}^{2}+2Y_{2}Y_{3})(Y_{1}^{3}+Y_{2}^{3}+Y_{3}^{2}-3Y_{1}Y_{2}Y_{3})
		\end{equation}
		\begin{equation}
			\label{7e34}
			Y_{(1')}^{10} = (Y_{3}^{2}+2Y_{1}Y_{2})(Y_{1}^{3}+Y_{2}^{3}+Y_{3}^{2}-3Y_{1}Y_{2}Y_{3})
		\end{equation}
		\begin{equation}
			\label{7e35}
			Y_{(3_{1})}^{10} = (Y_{1}^{2}+2Y_{2}Y_{3})^{2}\begin{pmatrix}
				Y_{1}\\
				Y_{2}\\
				Y_{3}
			\end{pmatrix}
		\end{equation}
		\begin{equation}
			\label{7e36}
			Y_{(3_{2})}^{10} = (Y_{3}^{2}+2Y_{1}Y_{2})^{2}\begin{pmatrix}
				Y_{2}\\
				Y_{3}\\
				Y_{1}
			\end{pmatrix}
		\end{equation}
		\begin{equation}
			\label{7e37}
			Y_{(3_{3})}^{10} = (Y_{1}^{2}+2Y_{2}Y_{3})(Y_{3}^{2}+2Y_{1}Y_{2})\begin{pmatrix}
				Y_{3}\\
				Y_{1}\\
				Y_{2}
			\end{pmatrix}
		\end{equation}
	\end{itemize}
	In the current work, we have considered weights $k=4,6,8$ and $10$ for realization of different textures in the neutrino mass matrix, as described below.\\
	
	\section{Modular weights and resulting neutrino mass matrices in doublet LRSM}\label{w75}
	The super-potential corresponding to modular symmetric LRSM with sterile neutrino will be given by,
	\begin{equation}
		\label{7e38}
		\mathcal{W} = l_{L}^{T}i\sigma_{2}\phi l_{R}^{C}Y+fYl_{R}^{T}i\sigma_{2}H_{R}S_{L}+\mu S_{L}S_{L}
	\end{equation}
	where $Y$ represents the modular Yukawa form, and the corresponding modular form will be assigned different weights namely, $k_{Y}=4,6,8$ and $10$ and as such each of the particle content of the model will have different modular weights. While assigning the weights, the condition that the sum of the modular weights in each term of the super-potential should be zero is being followed. Given below is a thorough discussion on different neutrino mass structures as a result of different weights assigned to modular forms.
	\begin{table}[!h]
		\centering
		\begin{tabular}{|c|c|c|c|c|c|c|}
			\hline
			Gauge group & $l_L$ & $l_R$ & $\phi$  & $H_R$ & $S_{L} $\\
			\hline
			$SU(3)_C$ & 1 & 1 & 1 & 1 & 1\\
			\hline
			$SU(2)_L$ & 2 & 1 & 2  & 1 & 1 \\
			\hline
			$SU(2)_R$ & 1 & 2 & 2 &  2 & 1 \\
			\hline
			$U(1)_{B-L}$ & -1 & -1 & 0 & -1 & 0 \\
			\hline
		\end{tabular}
		\caption{\centering Charge assignments for particle content of the model for LRSM gauge group.}
		\label{t1}
	\end{table}
	\subsection{For weight $k_{Y}=4$}
\begin{figure}[h]
		\centering
		\includegraphics[scale=0.45]{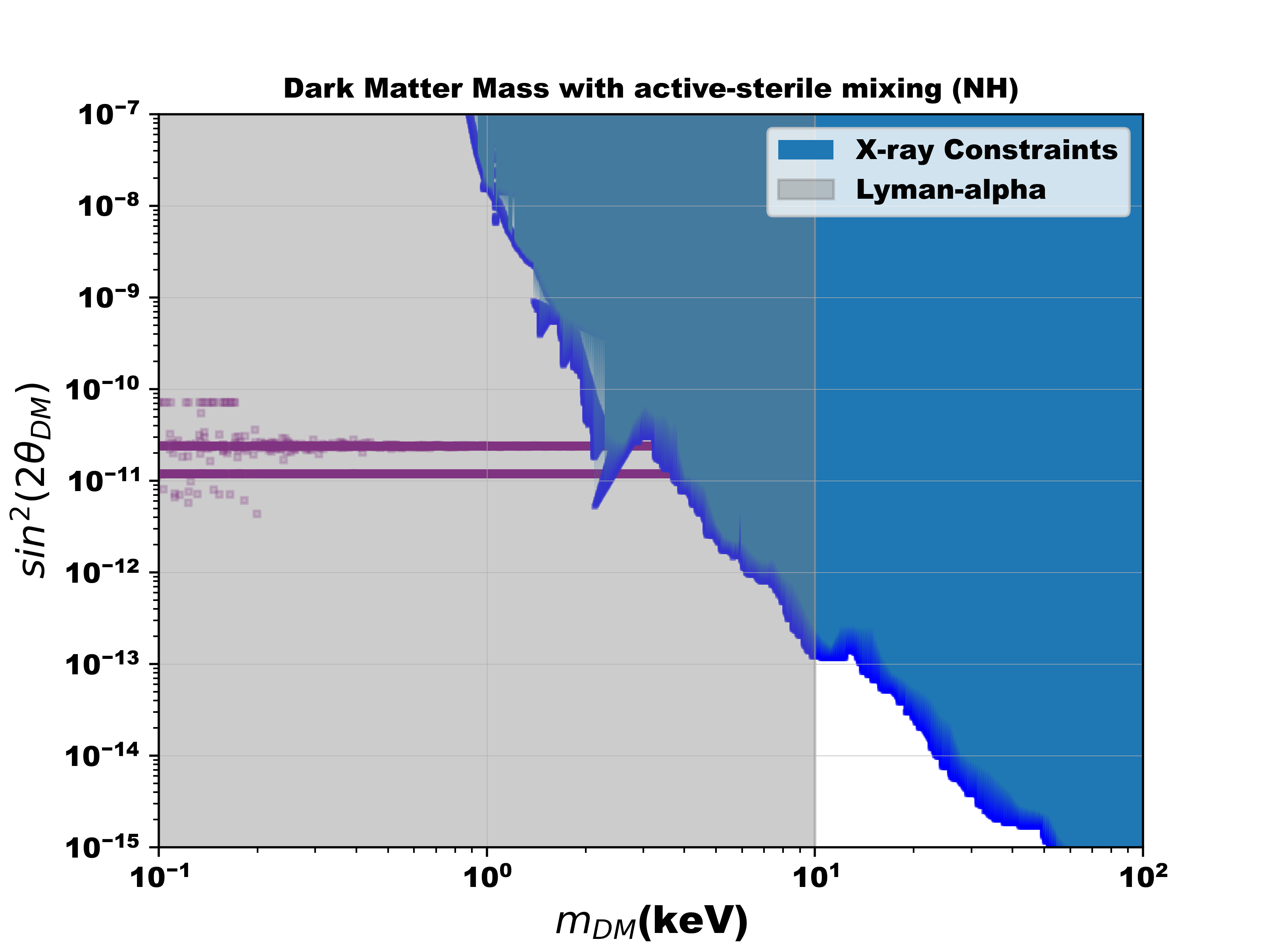}
		\includegraphics[scale=0.45]{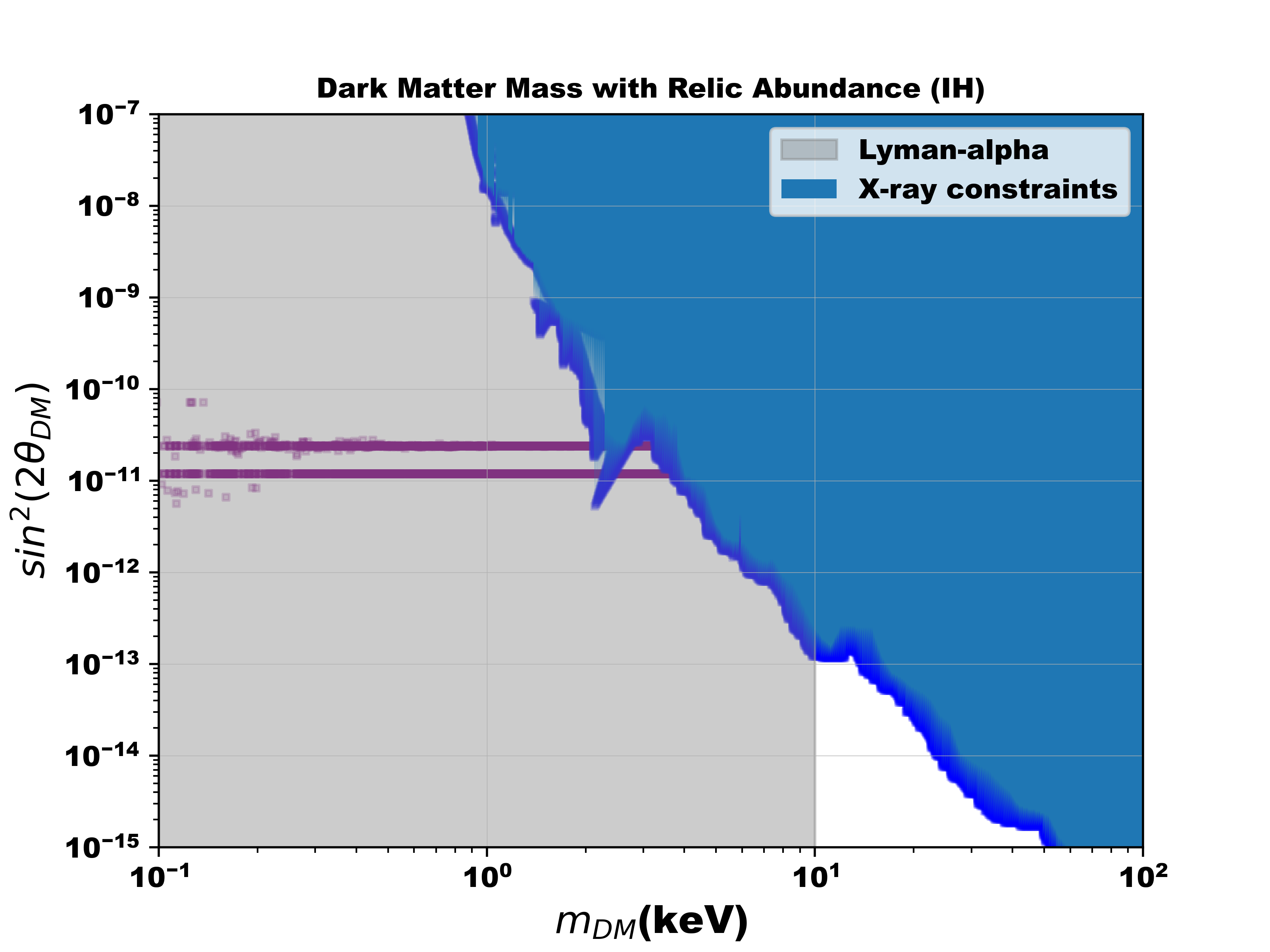}
		\caption{Dark matter mass as a function of active-sterile mixing where the gray region shows the constraints imposed by Lyman-$\alpha$ forest and blue region depicts the X-ray constraints. for $k_{Y}=4$.}
		\label{f41}
	\end{figure}

\begin{figure}[h]
		\centering
		\includegraphics[scale=0.45]{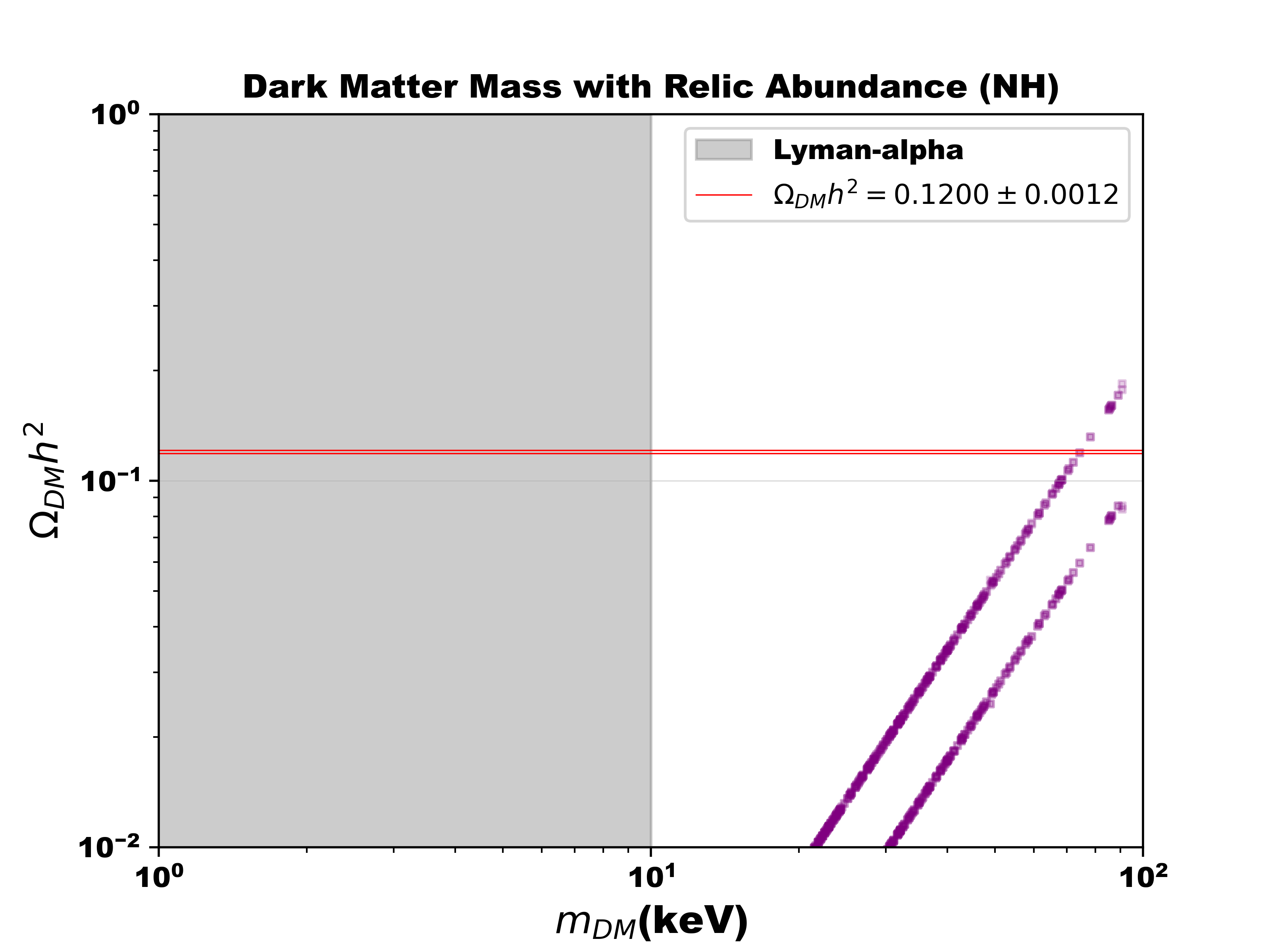}
		\includegraphics[scale=0.45]{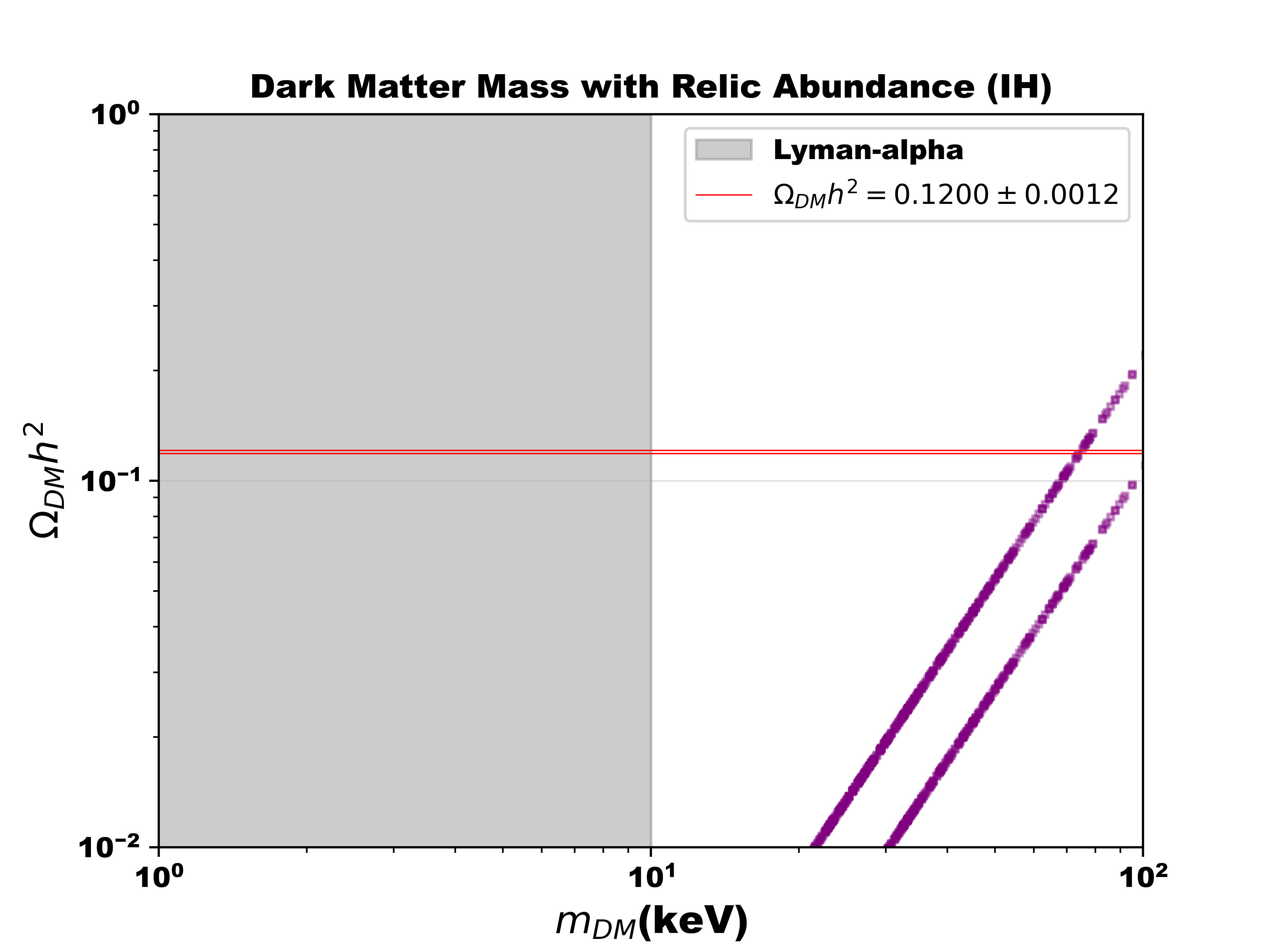}
		\caption{Variation of dark matter mass  relic abundance where the red band denotes the oberved value of relic abundance given by, $\Omega_{DM}h^{2}=0.1200\pm 0.0012$ for $k_{Y}=4$.}
		\label{f42}
	\end{figure}

\begin{figure}[h]
		\centering
		\includegraphics[scale=0.45]{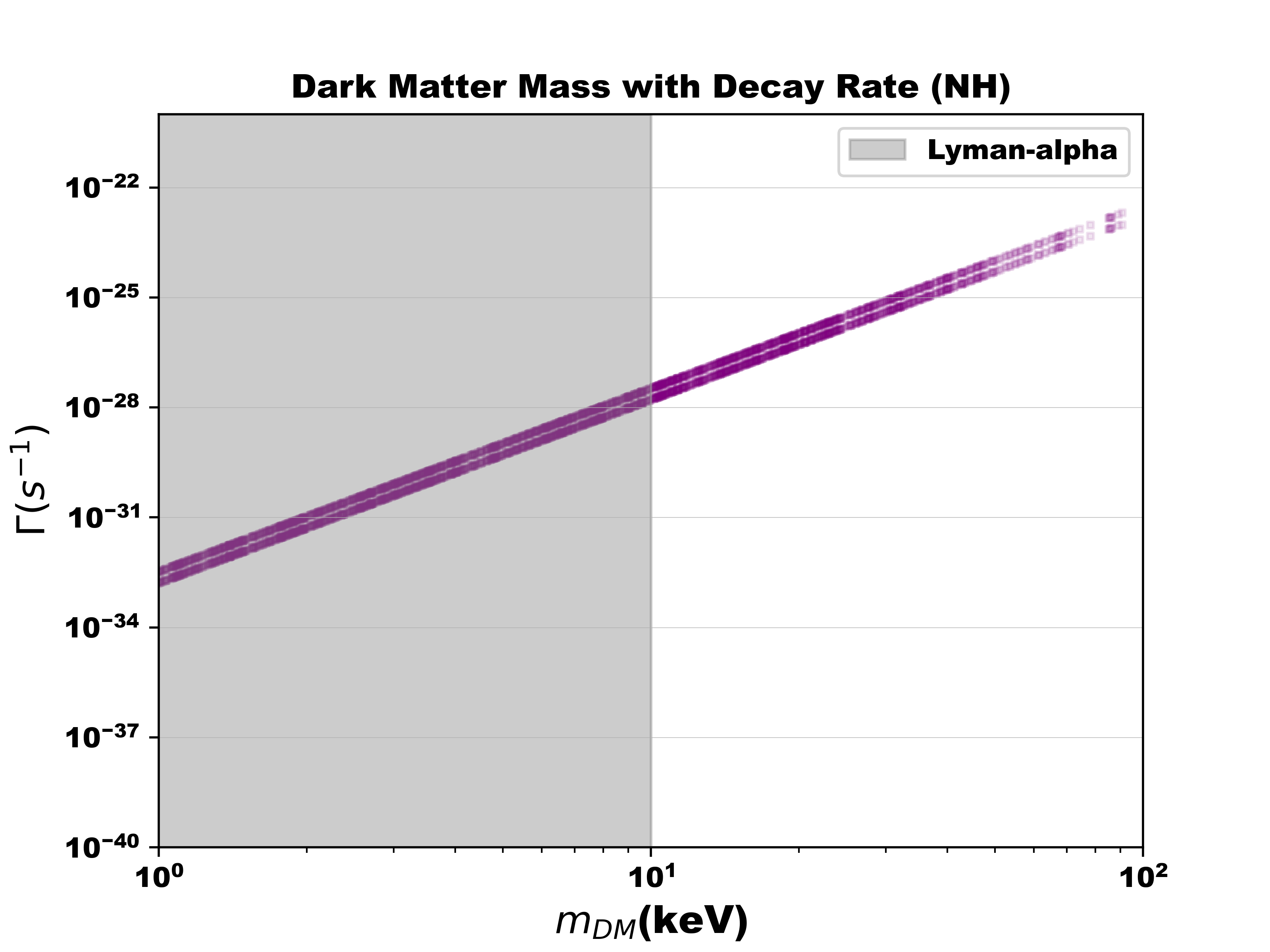}
		\includegraphics[scale=0.45]{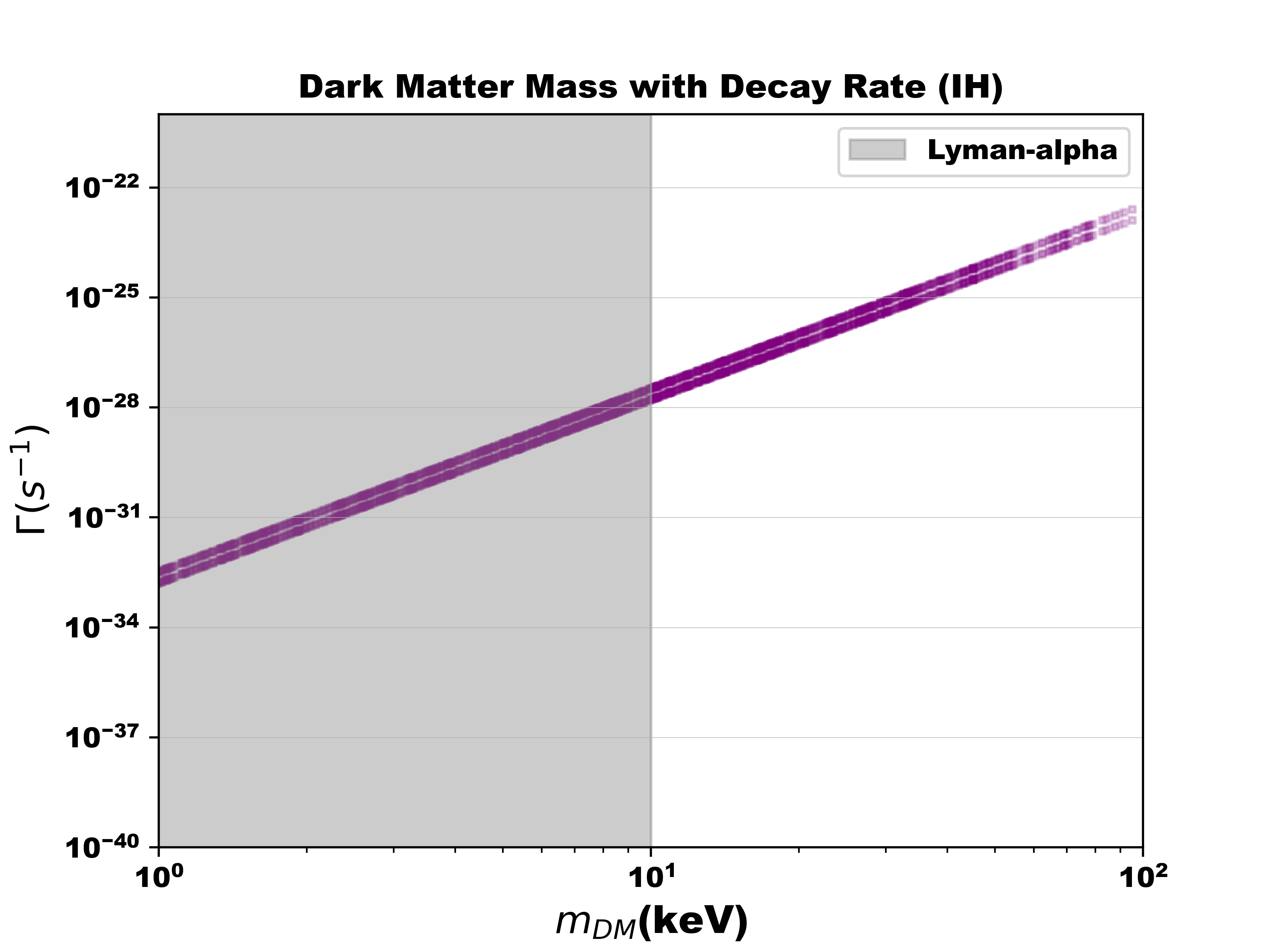}
		\caption{Variation of dark matter mass with decay rate of the dark matter candidate for weight $k_{Y}=4$.}
		\label{f43}
	\end{figure}

	\begin{table}[H]
		\begin{center}
			\begin{tabular}{|c|c|c|c|c|c|c|c|}
				\hline
				& $Y$ & $l_L$ & $l_R$ & $\phi$ & $H_{R}$ & $S_{L}$ \\
				\hline
				$A_{4}$ & 1,1' & 1,1',1" & 1,1',1" & 1 & 1 & 1\\
				\hline
				$k$ & 4 & -2 & -2 & -4 & 0 & -2\\
				\hline
			\end{tabular}
			\caption{\label{t2}Charge and weight assignments for the particle content under $A_4$ for $k_{Y}=4$.}
		\end{center}
	\end{table}	
	From the charge assignments quoted in table \ref{t2}, we obtain the respective mass matrices as,
	\begin{equation}
		\label{7e39}
		M_D =v\begin{pmatrix}
			Y_{1}^{4} & 0 & Y_{1'}^{4}\\
			0 & Y_{1'}^{4} & Y_{1}^{4}\\
			Y_{1'}^{4} & Y_{1}^{4} & 0
		\end{pmatrix}
	\end{equation}
	where, $v$ is VEV of the Higgs bidoublet $\phi$ and $v=\sqrt{k^{2}+k'^{2}} \simeq 246 GeV$, $k$ and $k'$ representing the VEVs of the neutral components of $\phi$. The active-sterile neutrino matrix is given by,
	\begin{equation}
		\label{7e40}
		M_{RS} =v_R f\begin{pmatrix}
			Y_{1}^{4} & 0 & Y_{1'}^{4}\\
			0 & Y_{1'}^{4} & Y_{1}^{4}\\
			Y_{1'}^{4} & Y_{1}^{4} & 0
		\end{pmatrix}
	\end{equation}
	where, $v_R$ is the VEV for the scalar doublet $H_R$ and $f$ is a dimensionless arbitrary parameter. The Majorana sterile neutrino mass matrix is given by,
	\begin{equation}
		\label{7e41}
		\mu = s \begin{pmatrix}
			Y_{1}^{4} & 0 & 0\\
			0 & 0 & Y_{1}^{4}\\
			0 & Y_{1}^{4} & 0
		\end{pmatrix}
	\end{equation}
	$s$ being an arbitrary parameter with dimensions of $keV$.\\
	The right-handed neutrino mass matrix obtained by using equation \eqref{7e10} is given as,
	\begin{equation}
		\label{7e42}
		M_{R}=\begin{pmatrix}
			-\frac{f^{2}v_{R}^{2}(Y_{1}^{2}+2Y_{2}Y_{3})}{s} & -\frac{f^{2}v_{R}^{2}(Y_{3}^{2}+2Y_{1}Y_{2})^{2}}{s(Y_{1}^{2}+2Y_{2}Y_{3})} & -\frac{2f^{2}v_{R}^{2}(Y_{3}^{2}+2Y_{1}Y_{2})}{s}\\
			-\frac{f^{2}v_{R}^{2}(Y_{3}^{2}+2Y_{1}Y_{2})^{2}}{s(Y_{1}^{2}+2Y_{2}Y_{3})} & -\frac{2f^{2}v_{R}^{2}(Y_{3}^{2}+2Y_{1}Y_{2})}{s} & -\frac{f^{2}v_{R}^{2}(Y_{1}^{2}+2Y_{2}Y_{3})}{s}\\
			-\frac{2f^{2}v_{R}^{2}(Y_{3}^{2}+2Y_{1}Y_{2})}{s} & -\frac{f^{2}v_{R}^{2}(Y_{1}^{2}+2Y_{2}Y_{3})}{s} & -\frac{f^{2}v_{R}^{2}(Y_{3}^{2}+2Y_{1}Y_{2})^{2}}{s(Y_{1}^{2}+2Y_{2}Y_{3})}
		\end{pmatrix}
	\end{equation}
	The light neutrino mass determined using double seesaw mechanism as given in the expression \eqref{7e9} turns out to be,
	\begin{equation}
		\label{7e43}
		M_\nu = \begin{pmatrix}
			\frac{s v^{2}Y_{2}(Y_2+2Y_3)}{f^{2}v_{R}^{2}} & 0 & 0\\
			0 & 0 & \frac{s v^{2}Y_{2}(Y_2+2Y_3)}{f^{2}v_{R}^{2}}\\
			0 & \frac{s v^{2}Y_{2}(Y_2+2Y_3)}{f^{2}v_{R}^{2}} & 0
		\end{pmatrix}
	\end{equation}
	It is evident from \eqref{7e43} that $M_{\nu}$ is symmetric in  nature. Although the matrix consists of more than two zeros, it has been considered to be allowed for the $3+1$ scenario which is under consideration. In figures \ref{f41} to \ref{f106}, NH refers to normal hierarchy and IH represents inverted hierarchy.
	It has been observed from figure \ref{f41} that the data points pertaining to the active-sterile mixing does not satisfy the constraints imposed by Lyman-$\alpha$ forest and X-ray constraints simultaneously and as such, a specific range of $m_{DM}$ satisfying both the constraints has not been achieved for $k_{Y}=4$.
	\\
Although, a specific set of range for $m_{DM}$ has not been observed, however, the data points clearly satisfy the limit imposed on the relic abundance of dark matter and also the decay rate has been found to lie between $10^{-37}$ to $10^{-22}$ which is almost negligible and is a satisfactory result.
\subsection{For weight $k_{Y}=6$}
	\begin{table}[H]
		\begin{center}
			\begin{tabular}{|c|c|c|c|c|c|c|c|}
				\hline
				& $Y$ & $l_L$ & $l_R$ & $\phi$ & $H_{R}$ & $S_{L}$ \\
				\hline
				$A_{4}$ & 1 & 1,1',1" & 1,1',1" & 1 & 1 & 1\\
				\hline
				$k$ & 6  & -3 & -3 & -6 & 0 & -3\\
				\hline
			\end{tabular}
			\caption{\label{t3}Charge and weight assignments for the particle content under $A_4$.}
		\end{center}
	\end{table}	
	From the charge assignments quoted in table \ref{t3}, we obtain the respective mass matrices as,
	\begin{equation}
		\label{7e44}
		M_D =v\begin{pmatrix}
			Y_{1}^{6} & 0 & 0\\
			0 & 0 & Y_{1}^{6}\\
			0 & Y_{1}^{6} & 0
		\end{pmatrix}
	\end{equation}
	The active-sterile neutrino matrix is given by,
	\begin{equation}
		\label{7e45}
		M_{RS} =v_R f\begin{pmatrix}
			Y_{1}^{6} & 0 & 0\\
			0 & 0 & Y_{1}^{6}\\
			0 & Y_{1}^{6} & 0
		\end{pmatrix}
	\end{equation}
	The Majorana sterile neutrino mass matrix is given by,
	\begin{equation}
		\label{7e46}
		\mu = s \begin{pmatrix}
			Y_{1}^{6} & 0 & 0\\
			0 & 0 & Y_{1}^{6}\\
			0 & Y_{1}^{6} & 0
		\end{pmatrix}
	\end{equation}
	The right-handed neutrino mass matrix is obtained as,
	\begin{equation}
		\label{7e47}
		M_{R}=\begin{pmatrix}
			-\frac{f^{2}v_{R}^{2}(Y_{1}^{3}+Y_{2}^{3}+Y_{3}^{3}-3Y_{1}Y_{2}Y_{3})}{s} & 0 & 0\\
			0 & 0 & -\frac{f^{2}v_{R}^{2}(Y_{1}^{3}+Y_{2}^{3}+Y_{3}^{3}-3Y_{1}Y_{2}Y_{3})}{s}\\
			0 & -\frac{f^{2}v_{R}^{2}(Y_{1}^{3}+Y_{2}^{3}+Y_{3}^{3}-3Y_{1}Y_{2}Y_{3})}{s} & 0
		\end{pmatrix}
	\end{equation}
	The light neutrino mass determined using double seesaw mechanism as given in the expression \eqref{7e9} turns out to be,
	\begin{equation}
		\label{7e48}
		M_\nu = \begin{pmatrix}
			\frac{s v^{2}(Y_1^{3}+Y_{2}^{3}+Y_{3}^{3}-3Y_{1}Y_{2}Y_{3})}{f^{2}v_{R}^{2}} & 0 & 0\\
			0 & 0 & \frac{s v^{2}(Y_1^{3}+Y_{2}^{3}+Y_{3}^{3}-3Y_{1}Y_{2}Y_{3})}{f^{2}v_{R}^{2}}\\
			0 & \frac{s v^{2}(Y_1^{3}+Y_{2}^{3}+Y_{3}^{3}-3Y_{1}Y_{2}Y_{3})}{f^{2}v_{R}^{2}} & 0
		\end{pmatrix}
	\end{equation}\\
	For weight $k_{Y}=6$, the resulting mass matrix has been derived, however the results were not satisfactorily obtained for the same and as such they have not been presented in the manuscript.
	
	\subsection{For weight $k_{Y}=8$}
	\begin{table}[H]
		\begin{center}
			\begin{tabular}{|c|c|c|c|c|c|c|c|}
				\hline
				& $Y$ & $l_L$ & $l_R$ & $\phi$ & $H_{R}$ & $S_{L}$ \\
				\hline
				$A_{4}$ & 1,1',1" & 1,1',1" & 1,1',1" & 1 & 1 & 1\\
				\hline
				$k$ & 8  & -4 & -4 & -8 & 0 & -4\\
				\hline
			\end{tabular}
			\caption{\label{t4}Charge and weight assignments for the particle content under $A_4$.}
		\end{center}
	\end{table}	
	From the charge assignments quoted in table \ref{t4}, we obtain the respective mass matrices as,
	\begin{equation}
		\label{7e49}
		M_D =v\begin{pmatrix}
			Y_{1}^{8} & Y_{1^"}^{8} & Y_{1'}^{8}\\
			Y_{1^"}^{8} & Y_{1'}^{8} & Y_{1}^{8}\\
			Y_{1'}^{8} & Y_{1}^{8} & Y_{1^"}^{8}
		\end{pmatrix}
	\end{equation}
	The active-sterile neutrino matrix is given by,
	\begin{equation}
		\label{7e50}
		M_{RS} =v_R f\begin{pmatrix}
			Y_{1}^{8} & Y_{1^"}^{8} & Y_{1'}^{8}\\
			Y_{1^"}^{8} & Y_{1'}^{8} & Y_{1}^{8}\\
			Y_{1'}^{8} & Y_{1}^{8} & Y_{1^"}^{8}
		\end{pmatrix}
	\end{equation}
	The Majorana sterile neutrino mass matrix is given by,
	\begin{equation}
		\label{7e51}
		\mu = s \begin{pmatrix}
			Y_{1}^{8} & 0 & 0\\
			0 & 0 & Y_{1}^{8}\\
			0 & Y_{1}^{8} & 0
		\end{pmatrix}
	\end{equation}
	$M_{R}$ for weight $k_{Y}=8$ is obtained as,
	\begin{equation}
		\label{7e52}
		\scriptsize M_{R}=a\begin{pmatrix}
			-\frac{(-(Y_{1}^{2}+Y_{2}^{3})^{3}-2Y_{3}^{2}(2Y_{1}+Y_{3})^{2}(2Y_{1}Y_{2}+Y_{3}^{2}))}{s(Y_{1}^{2}+2Y_{2}Y_{3})} & -\frac{3Y_{3}^{4}+4Y_{1}Y_{3}^{2}(Y_{2}+2Y_{3})+4Y_{1}^{2}(Y_{2}^{2}+2Y_{3}^{2})}{s} & \frac{-Y_{3}^{4}(2Y_{1}+Y_{3})^{4}-2(Y_{1}^{2}+2Y_{2}Y_{3})^{3}(2Y_{1}Y_{2}+Y_{3}^{2})}{s(Y_{1}^{2}+2Y_{2}Y_{3})^{2}}\\
			-\frac{(3Y_{3}^{4}+4Y_{1}Y_{3}^{2})(Y_{2}+2Y_{3})+4Y_{1}^{2}(Y_{2}^{2}+2Y_{3}^{2})}{s} & \frac{(-Y_{3}^{4}(2Y_{1}+Y_{3})^{4}-2(Y_{1}^{2}+2Y_{2}Y_{3})^{3}(2Y_{1}Y_{2}+Y_{3}^{2})}{s(Y_{1}^{2}+2Y_{2}Y_{3})^{2}} &  -\frac{(-(Y_{1}^{2}+Y_{2}^{3})^{3}-2Y_{3}^{2}(2Y_{1}+Y_{3})^{2}(2Y_{1}Y_{2}+Y_{3}^{2}))}{s(Y_{1}^{2}+2Y_{2}Y_{3})} \\
			\frac{(-Y_{3}^{4}(2Y_{1}+Y_{3})^{4}-2(Y_{1}^{2}+2Y_{2}Y_{3})^{3}(2Y_{1}Y_{2}+Y_{3}^{2})}{s(Y_{1}^{2}+2Y_{2}Y_{3})^{2}} & -\frac{(-(Y_{1}^{2}+Y_{2}^{3})^{3}-2Y_{3}^{2}(2Y_{1}+Y_{3})^{2}(2Y_{1}Y_{2}+Y_{3}^{2}))}{s(Y_{1}^{2}+2Y_{2}Y_{3})} & -\frac{(3Y_{3}^{4}+4Y_{1}Y_{3}^{2}(Y_{2}+2Y_{3})+4Y_{1}^{2}(Y_{2}^{2}+2Y_{3}^{2})}{s} 
		\end{pmatrix}
	\end{equation}
	where, $a=f^{2}v_{R}^{2}$

\begin{figure}[h]
		\centering
		\includegraphics[scale=0.45]{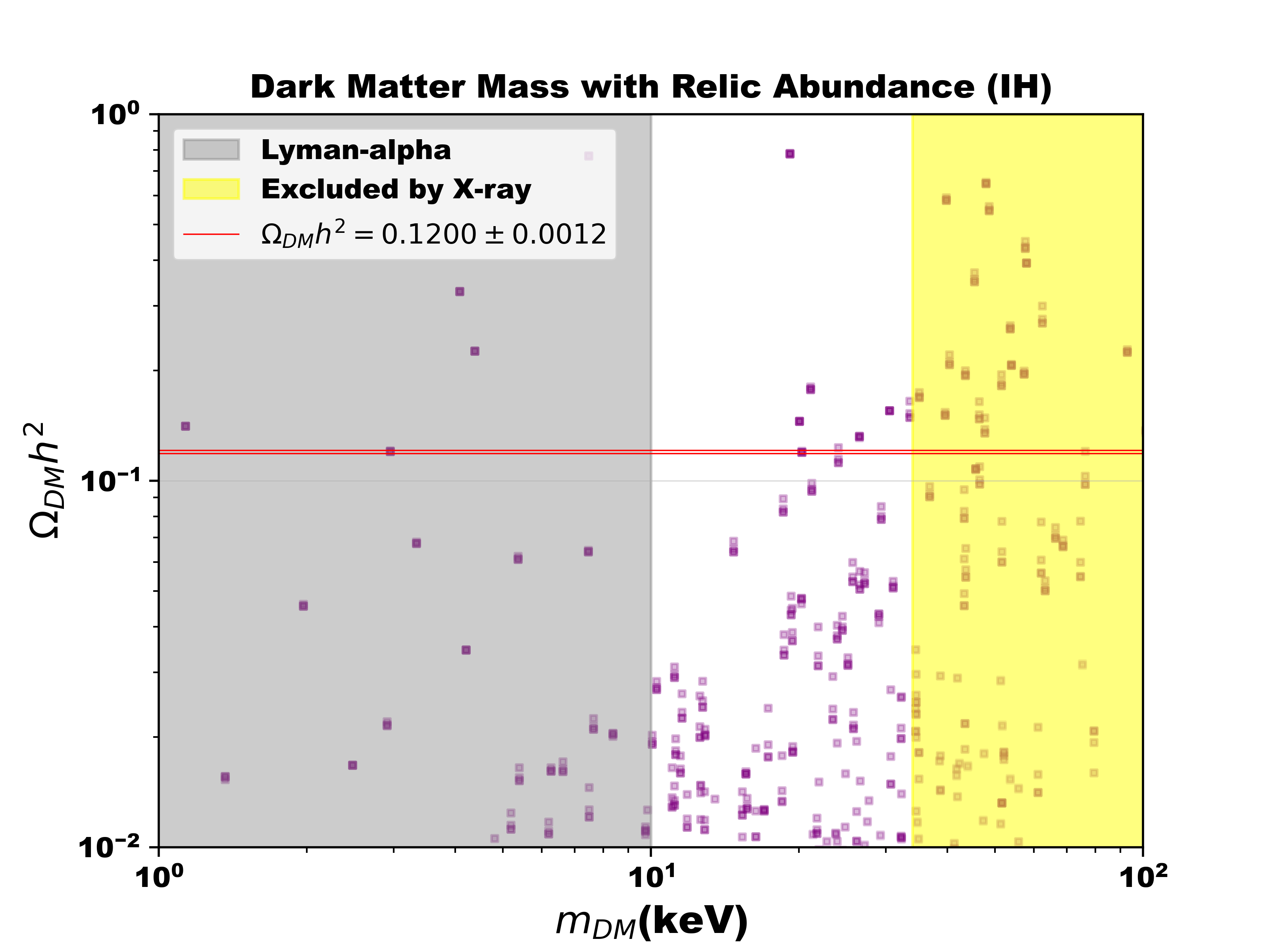}
		\includegraphics[scale=0.45]{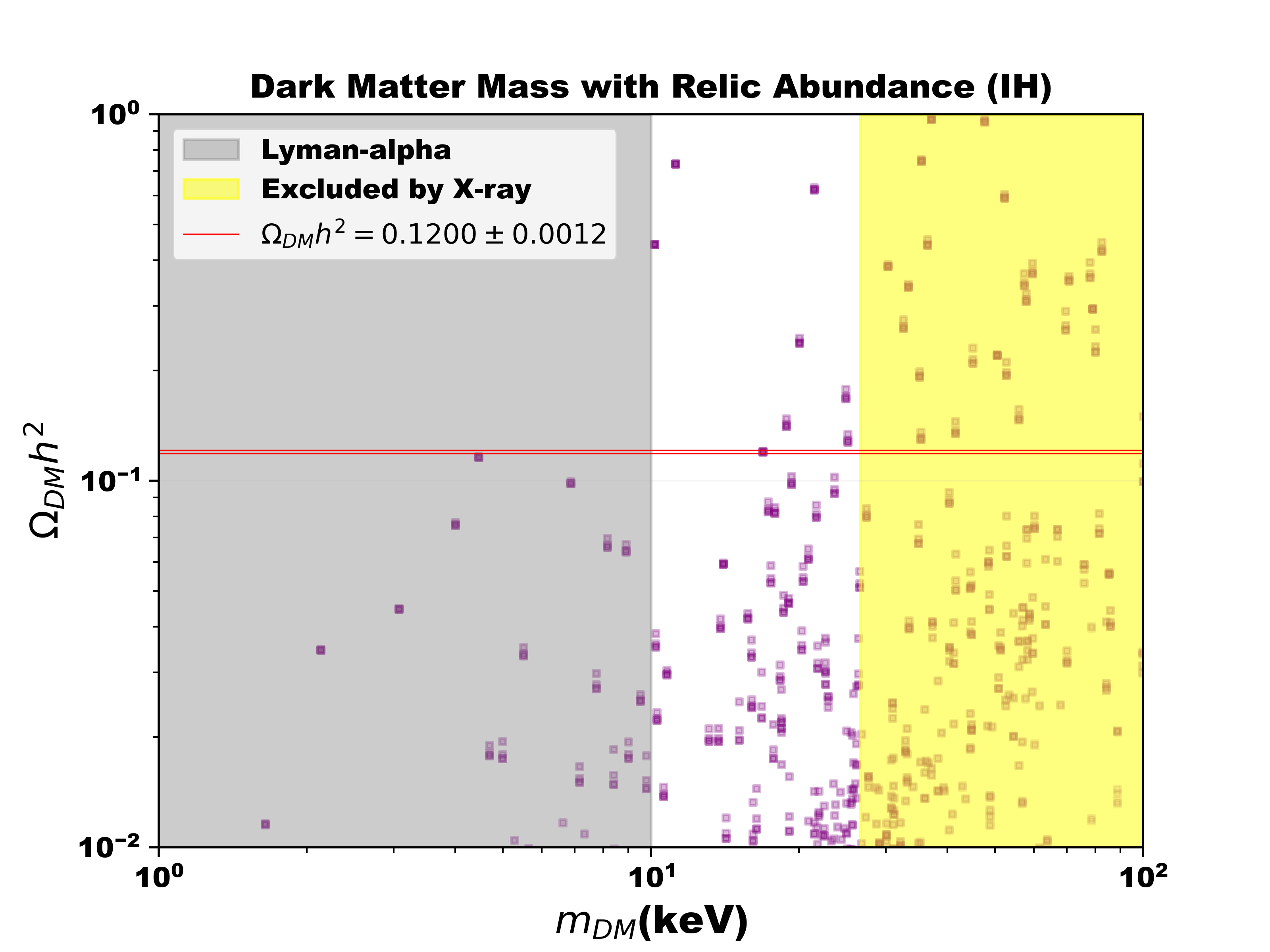}
		\caption{Variation of dark matter mass  relic abundance where the red band denotes the oberved value of relic abundance given by, $\Omega_{DM}h^{2}=0.1200\pm 0.0012$ for $k_{Y}=8$.}
		\label{f82}
	\end{figure}

\begin{figure}[h]
		\centering
		\includegraphics[scale=0.45]{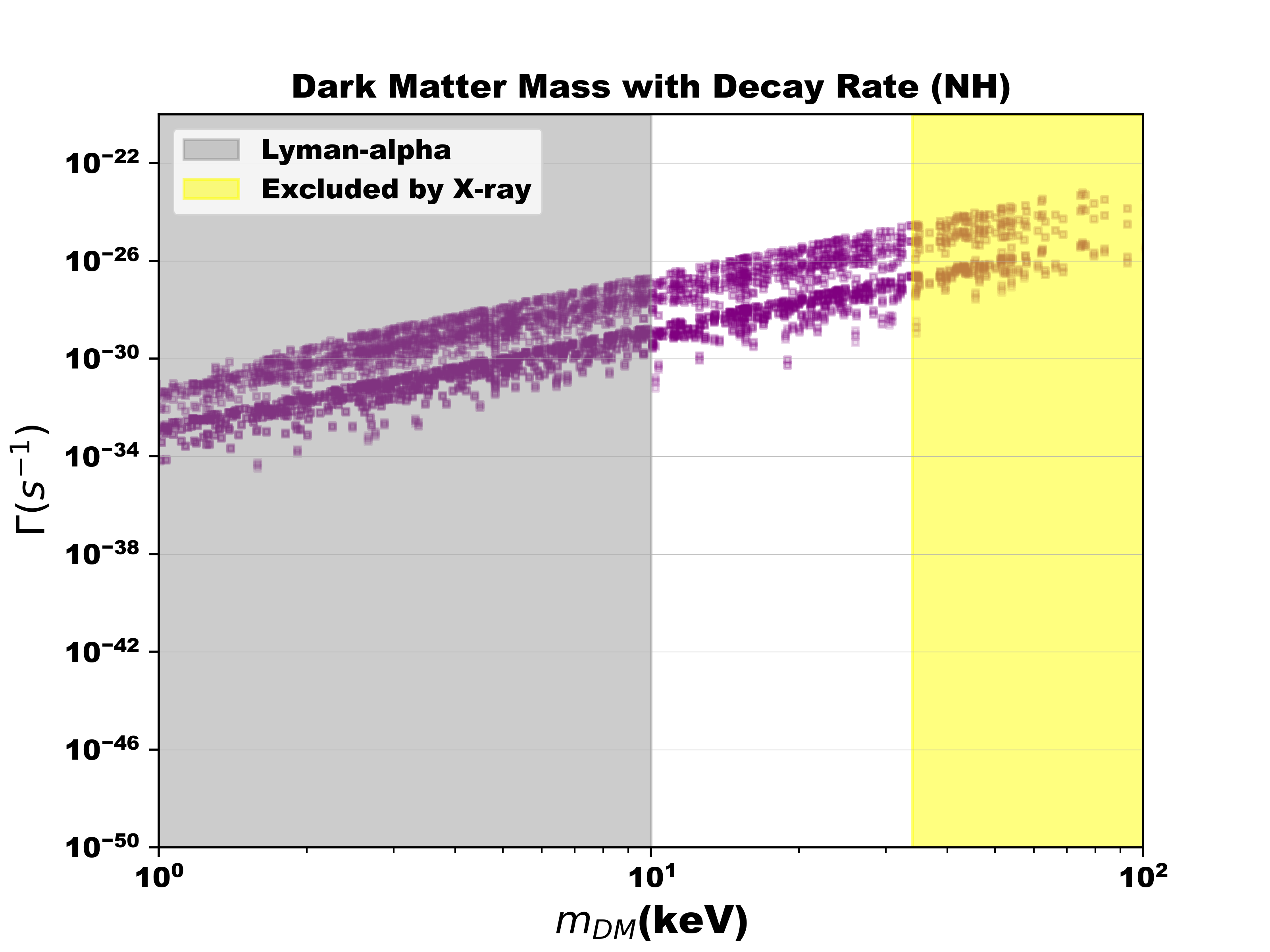}
		\includegraphics[scale=0.45]{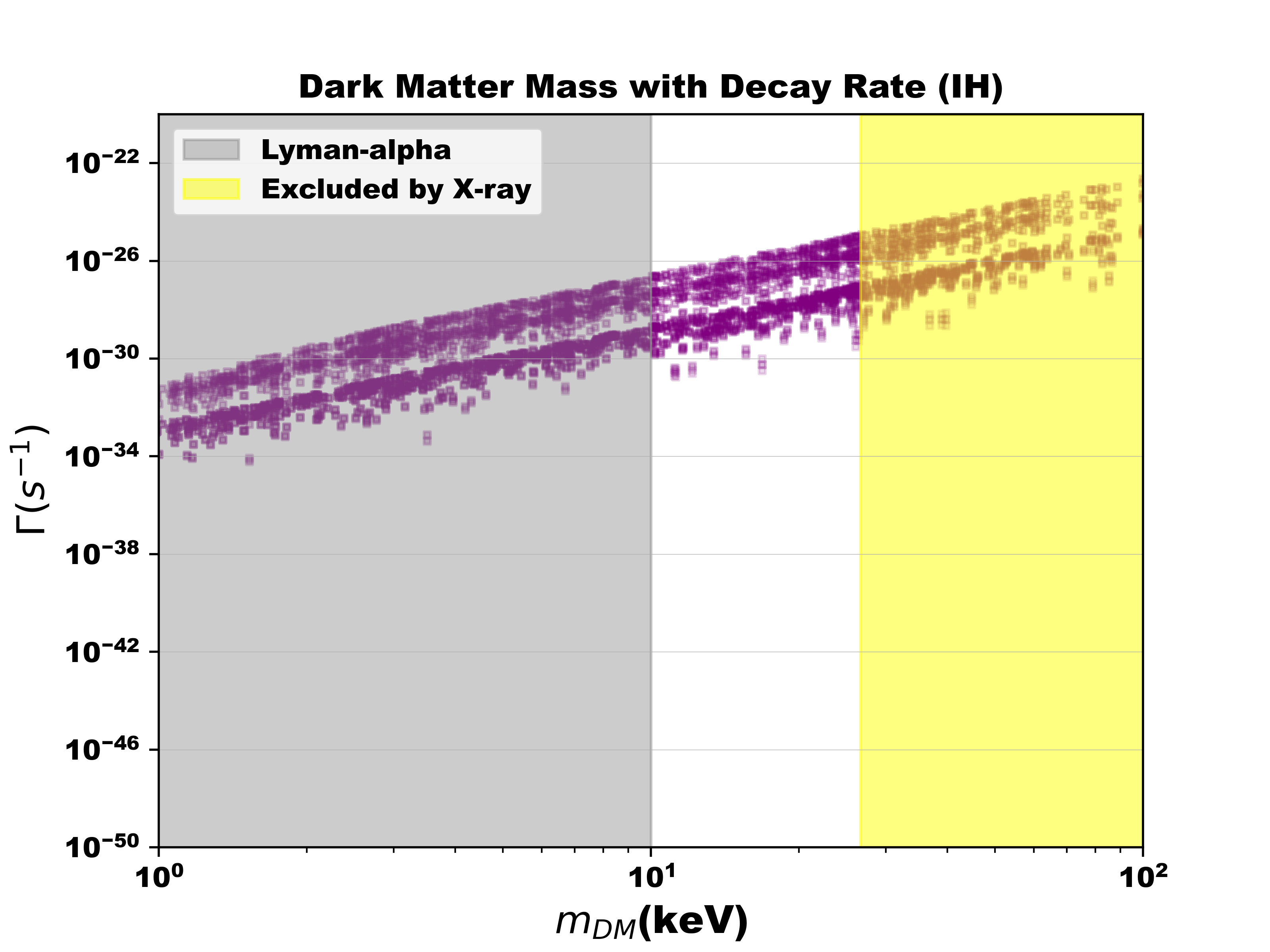}
		\caption{Variation of dark matter mass with decay rate of the dark matter candidate for weight $k_{Y}=8$.}
		\label{f83}
	\end{figure}

	The light neutrino mass determined using double seesaw mechanism as given in the expression \eqref{7e9} turns out to be,
	\begin{equation}
		\label{7e53}
		M_\nu = \begin{pmatrix}
			\frac{s v^{2}(Y_{1}^{2}+2Y_{2}Y_{3})}{f^{2}v_{R}^{2}} & 0 & 0\\
			0 & 0 & \frac{s v^{2}(Y_{1}^{2}+2Y_{2}Y_{3})}{f^{2}v_{R}^{2}}\\
			0 & \frac{s v^{2}(Y_{1}^{2}+2Y_{2}Y_{3})}{f^{2}v_{R}^{2}} & 0
		\end{pmatrix}
	\end{equation}
		\begin{figure}[h]
		\centering
		\includegraphics[scale=0.45]{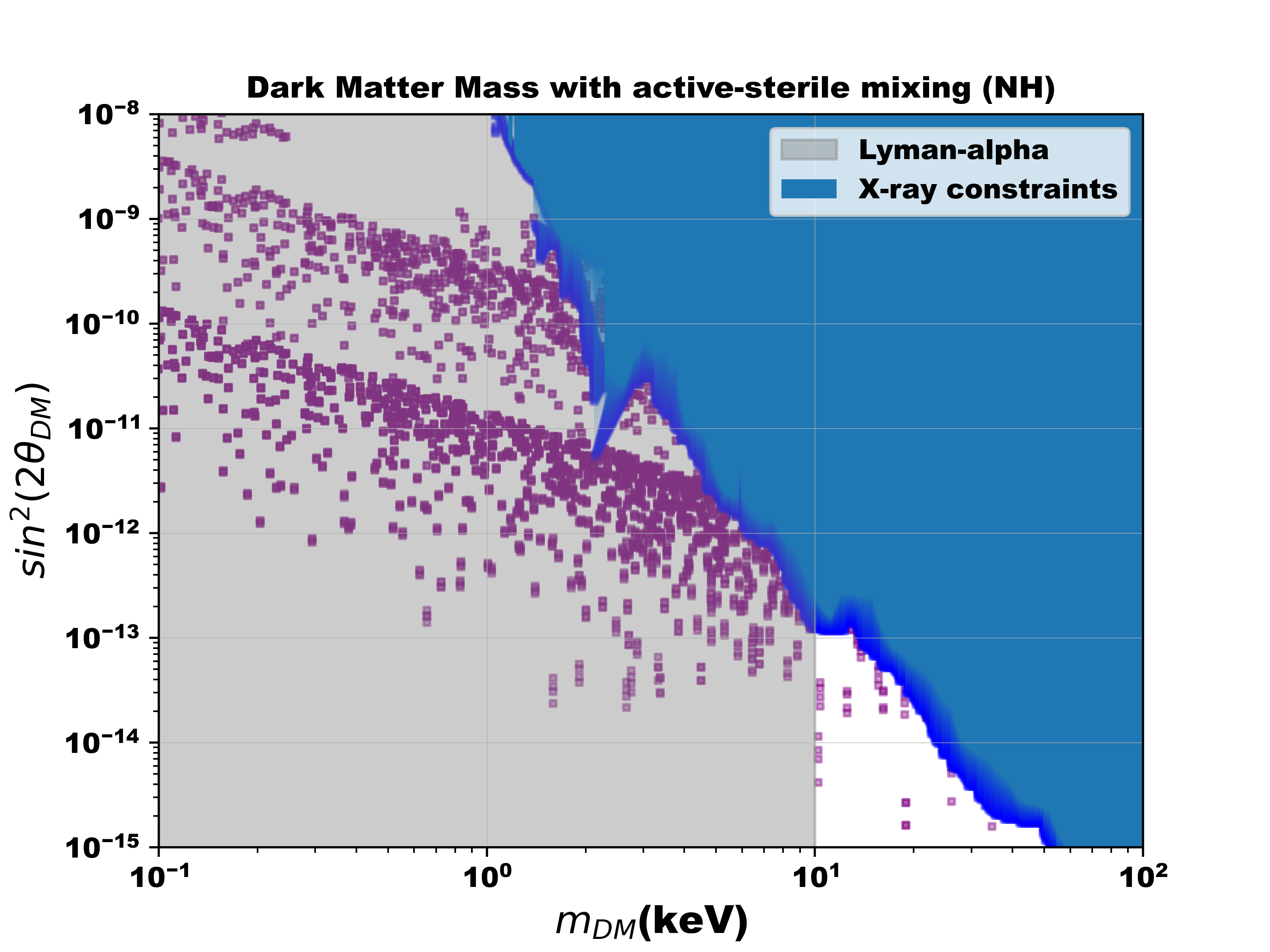}
		\includegraphics[scale=0.45]{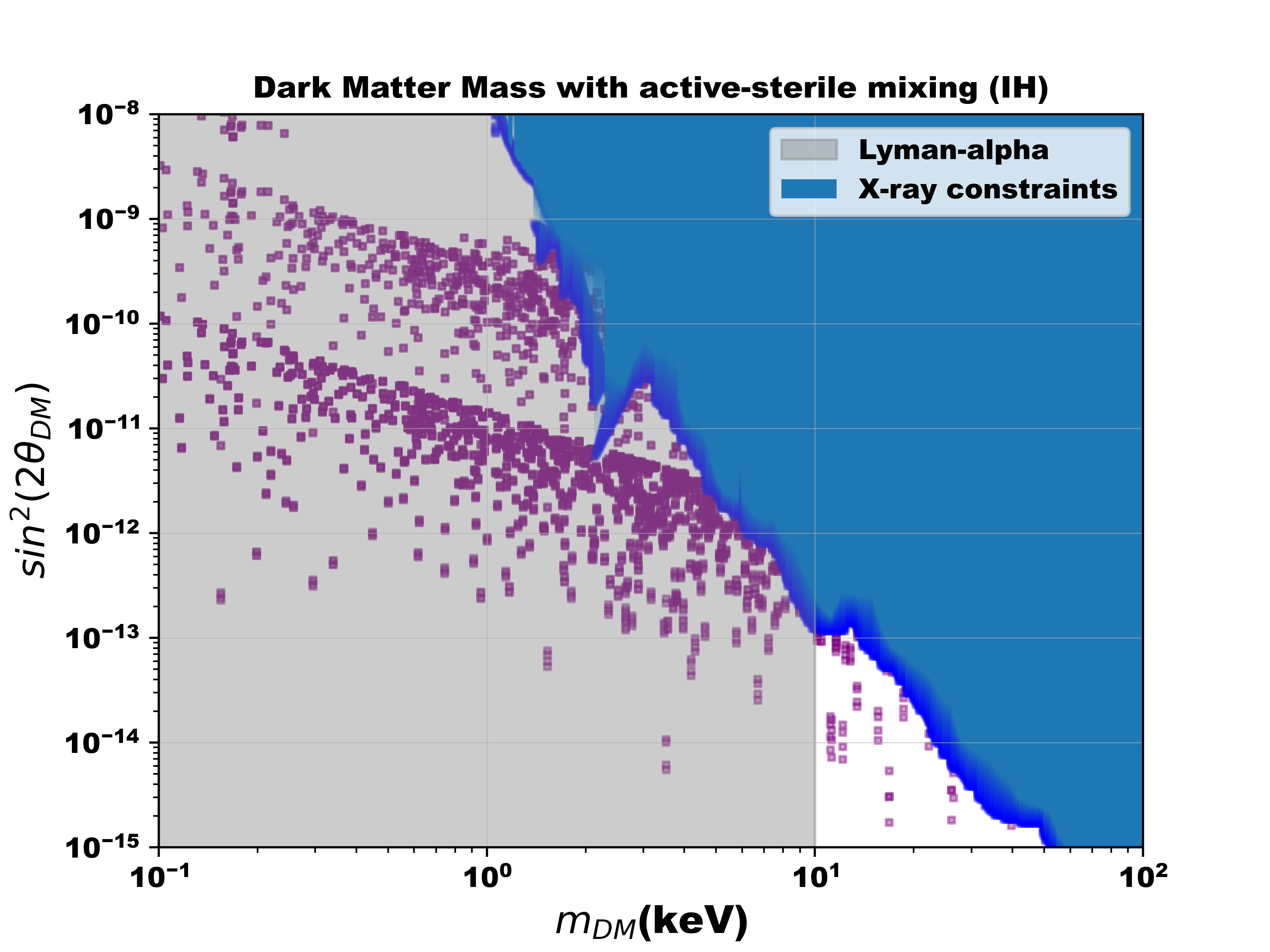}
		\caption{Dark matter mass as a function of active-sterile mixing where the gray region shows the constraints imposed by Lyman-$\alpha$ forest and blue region depicts the X-ray constraints. for $k_{Y}=8$.}
		\label{f81}
	\end{figure}\\
	From figure \ref{f81}, it has been observed that for normal hierarchy, the dark matter mass satisfying the Lyman- $\alpha$ and X-ray constraints ranges from 10 keV to 33.98 keV and for inverted hierarchy it falls in between 10 keV to 26.632 keV. In figures \ref{f82} and \ref{f83}, this range has been taken into consideration.
	
	\subsection{For weight $k_{Y}=10$}
	\begin{table}[H]
		\begin{center}
			\begin{tabular}{|c|c|c|c|c|c|c|c|}
				\hline
				& $Y$ & $l_L$ & $l_R$ & $\phi$ & $H_{R}$ & $S_{L}$ \\
				\hline
				$A_{4}$ & 1,1' & 1,1',1" & 1,1',1" & 1 & 1 & 1\\
				\hline
				$k$ & 10  & -5 & -5 & -10 & 0 & -5\\
				\hline
			\end{tabular}
			\caption{\label{t5}Charge and weight assignments for the particle content under $A_4$.}
		\end{center}
	\end{table}	
	From the charge assignments quoted in table \ref{t5}, we obtain the respective mass matrices as,
	\begin{equation}
		\label{7e54}
		M_D =v\begin{pmatrix}
			Y_{1}^{10} & 0 & Y_{1'}^{10}\\
			0 & Y_{1'}^{10} & Y_{1}^{10}\\
			Y_{1'}^{10} & Y_{1}^{10} & 0
		\end{pmatrix}
	\end{equation}
	The active-sterile neutrino matrix is given by,
	\begin{equation}
		\label{7e55}
		M_{RS} =v_R f\begin{pmatrix}
			Y_{1}^{10} & 0 & Y_{1'}^{10}\\
			0 & Y_{1'}^{10} & Y_{1}^{10}\\
			Y_{1'}^{10} & Y_{1}^{10} & 0
		\end{pmatrix}
	\end{equation}
	The Majorana sterile neutrino mass matrix is given by,
	\begin{equation}
		\label{7e56}
		\mu = s \begin{pmatrix}
			Y_{1}^{10} & 0 & 0\\
			0 & 0 & Y_{1}^{10}\\
			0 & Y_{1}^{10} & 0
		\end{pmatrix}
	\end{equation}
	And the right-handed neutrino mass matrix is given by,
	\begin{equation}
		\label{7e57}
		\scriptsize M_{R}=\begin{pmatrix}
			-\frac{f^2 v_R^2 (Y_1^2 + 2 Y_2 Y_3) (Y_1^3 + Y_2^3 - 3 Y_1 Y_2 Y_3 + Y_3^3)}{s} & -\frac{f^2 v_R^2 (2 Y_1 Y_2 + Y_3^2)^2 (Y_1^3 + Y_2^3 - 3 Y_1 Y_2 Y_3 + Y_3^3)}{s(Y_1^2 + 2 Y_2 Y_3)} & -\frac{2 f^2 v_R^2 (2 Y_1 Y_2 + Y_3^2) (Y_1^3 + Y_2^3 - 3 Y_1 Y_2 Y_3 + Y_3^3)}{s}\\
			-\frac{f^2 v_R^2 (2 Y_1 Y_2 + Y_3^2)^2 (Y_1^3 + Y_2^3 - 3 Y_1 Y_2 Y_3 + Y_3^3)}{s(Y_1^2 + 2 Y_2 Y_3)} & -\frac{2 f^2 v_R^2 (2 Y_1 Y_2 + Y_3^2) (Y_1^3 + Y_2^3 - 3 Y_1 Y_2 Y_3 + Y_3^3)}{s} &  -\frac{f^2 v_R^2 (Y_1^2 + 2 Y_2 Y_3) (Y_1^3 + Y_2^3 - 3 Y_1 Y_2 Y_3 + Y_3^3)}{s} \\
			-\frac{2 f^2 v_R^2 (2 Y_1 Y_2 + Y_3^2) (Y_1^3 + Y_2^3 - 3 Y_1 Y_2 Y_3 + Y_3^3)}{s} & -\frac{f^2 v_R^2 (Y_1^2 + 2 Y_2 Y_3) (Y_1^3 + Y_2^3 - 3 Y_1 Y_2 Y_3 + Y_3^3)}{s}& -\frac{f^2 v_R^2 (2 Y_1 Y_2 + Y_3^2)^2 (Y_1^3 + Y_2^3 - 3 Y_1 Y_2 Y_3 + Y_3^3)}{s(Y_1^2 + 2 Y_2 Y_3)} 
		\end{pmatrix}
	\end{equation}  
	
	The light neutrino mass determined using double seesaw mechanism as given in the expression \eqref{7e9} turns out to be,
	\begin{equation}
		\label{7e58}
		\scriptsize M_\nu = \begin{pmatrix}
			\frac{s v^{2}(Y_{1}^{2}+2Y_{2}Y_{3})(Y_{1}^{3}+Y_{2}^{3}+Y_{3}^{3}-3Y_{1}Y_{2}Y_{3})}{f^{2}v_{R}^{2}} & 0 & 0\\
			0 & 0 & \frac{s v^{2}(Y_{1}^{2}+2Y_{2}Y_{3})(Y_{1}^{3}+Y_{2}^{3}+Y_{3}^{3}-3Y_{1}Y_{2}Y_{3})}{f^{2}v_{R}^{2}}\\
			0 & \frac{s v^{2}(Y_{1}^{2}+2Y_{2}Y_{3})(Y_{1}^{3}+Y_{2}^{3}+Y_{3}^{3}-3Y_{1}Y_{2}Y_{3})}{f^{2}v_{R}^{2}} & 0
		\end{pmatrix}
	\end{equation}
	From equations \eqref{7e43},\eqref{7e48},\eqref{7e53} and \eqref{7e58}, it has been observed that all the weights correspond to the same structure of the resulting neutrino mass, but what is more important is the structure of the active-sterile mixing matrix $M_{RS}$ and also the mass matrix for right-handed neutrinos ($M_{R}$) for each weight. $M_{RS}$ is important for the determination of active sterile mixing angle $\theta_{DM}$ and to determine the allowed region for the mass of the DM candidate (sterile neutrino) and $M_{R}$ is necessary for the study of leptogenesis within the model as has been described in the succeeding sections.
		\begin{figure}[h]
		\centering
		\includegraphics[scale=0.45]{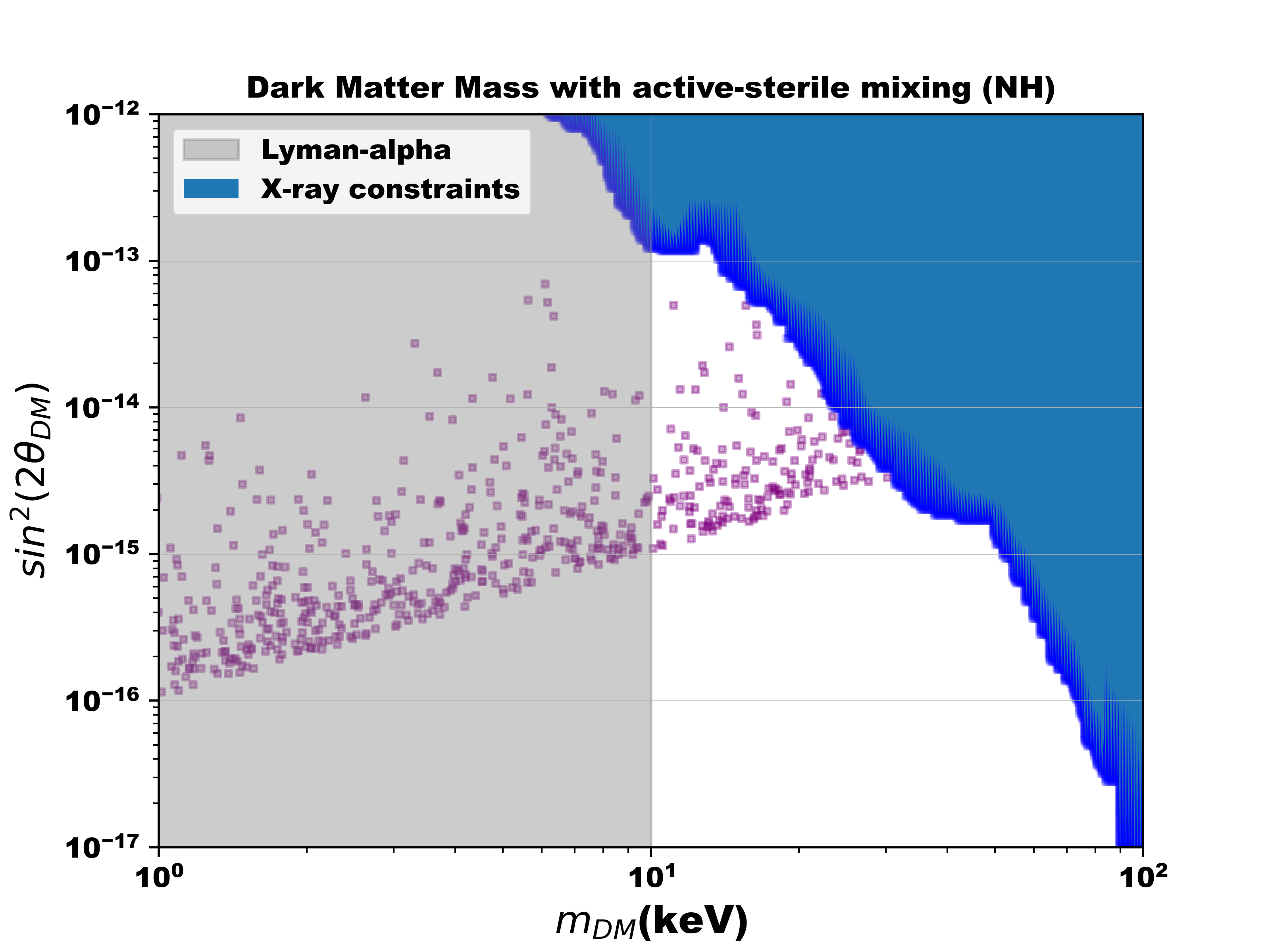}
		\includegraphics[scale=0.45]{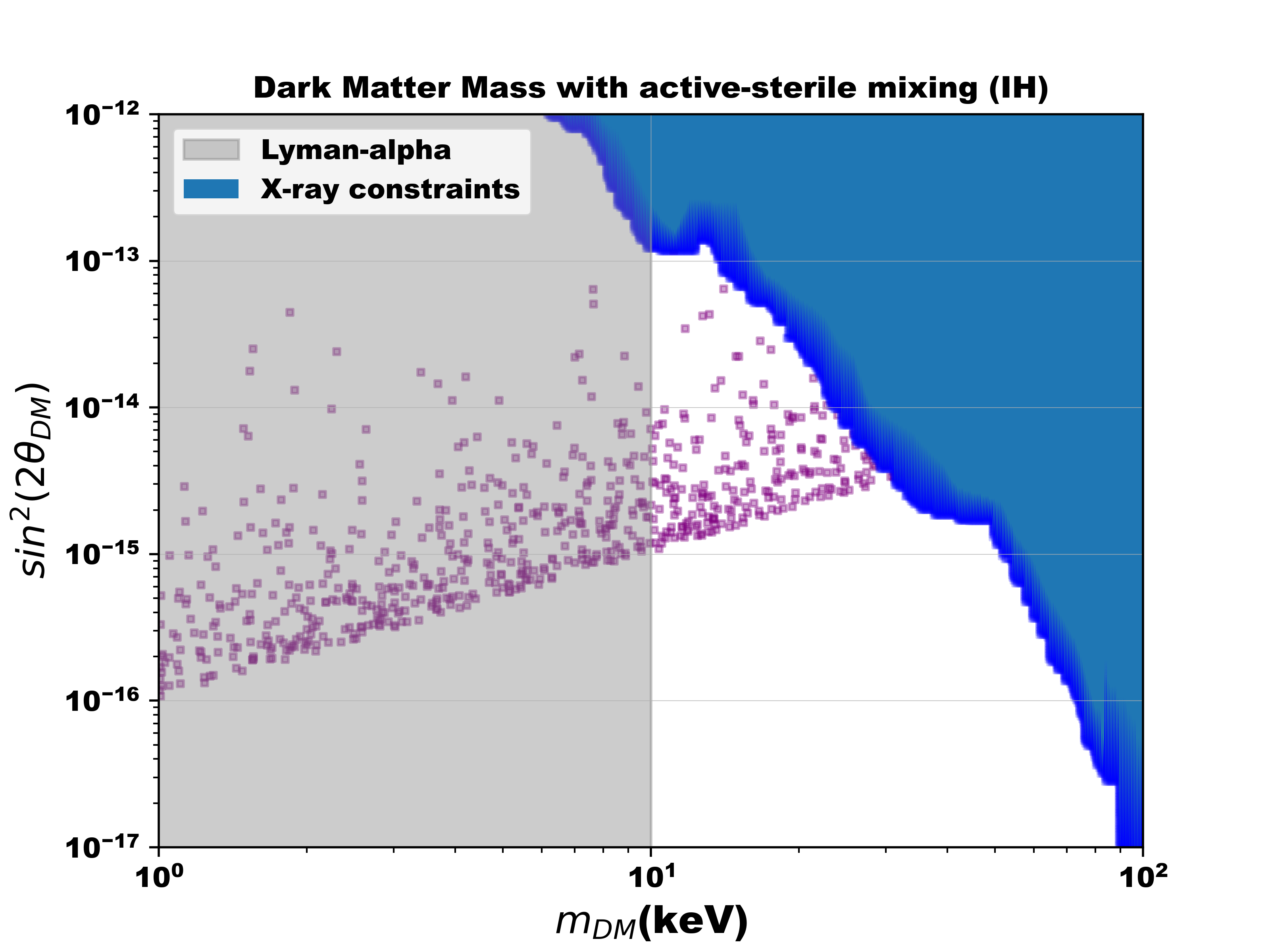}
		\caption{Dark matter mass as a function of active-sterile mixing where the gray region shows the constraints imposed by Lyman-$\alpha$ forest and blue region depicts the X-ray constraints. for $k_{Y}=10$.}
		\label{f101}
	\end{figure}\\
	From figure \ref{f101}, it can be predicted that the allowed mass range for normal hierarchy falls between 10 keV to 30.39 keV and for inverted hierarchy it ranges from 10 keV to 28.1577 keV.
	\begin{figure}[h]
		\centering
		\includegraphics[scale=0.45]{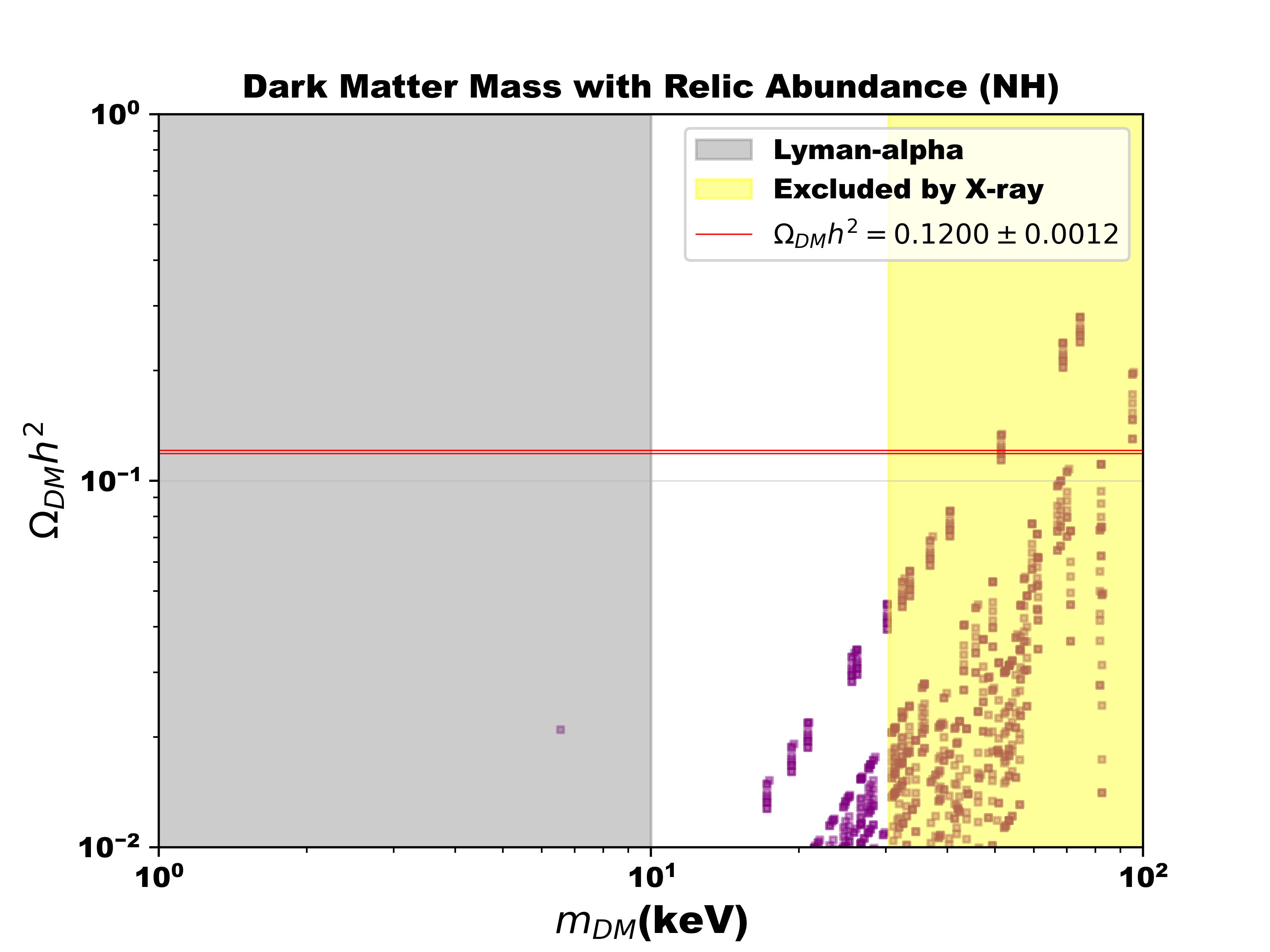}
		\includegraphics[scale=0.45]{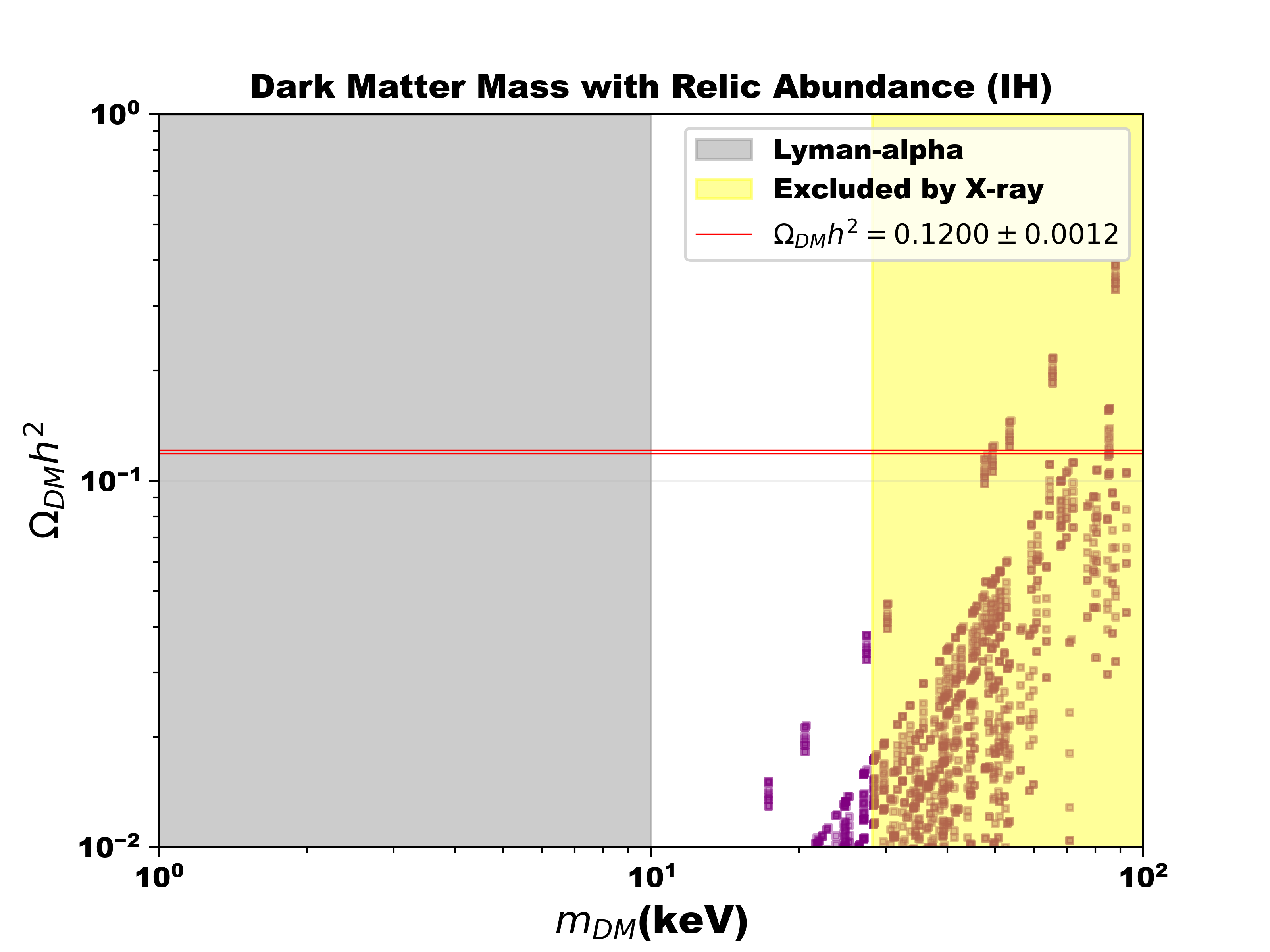}
		\caption{Variation of dark matter mass  relic abundance where the red band denotes the oberved value of relic abundance given by, $\Omega_{DM}h^{2}=0.1200\pm 0.0012$ for $k_{Y}=10$.}
		\label{f102}
	\end{figure}
	\begin{figure}[h]
		\centering
		\includegraphics[scale=0.45]{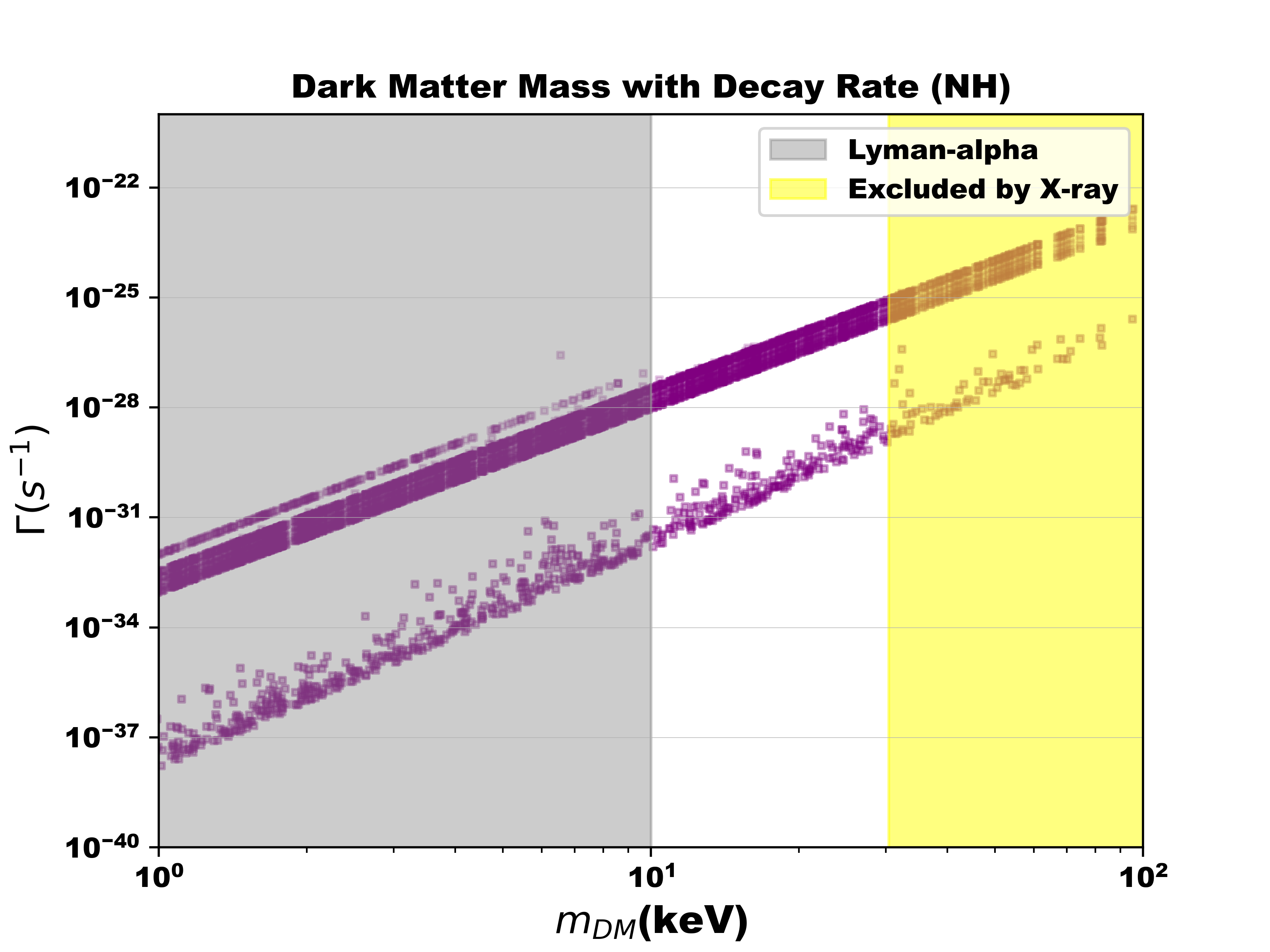}
		\includegraphics[scale=0.45]{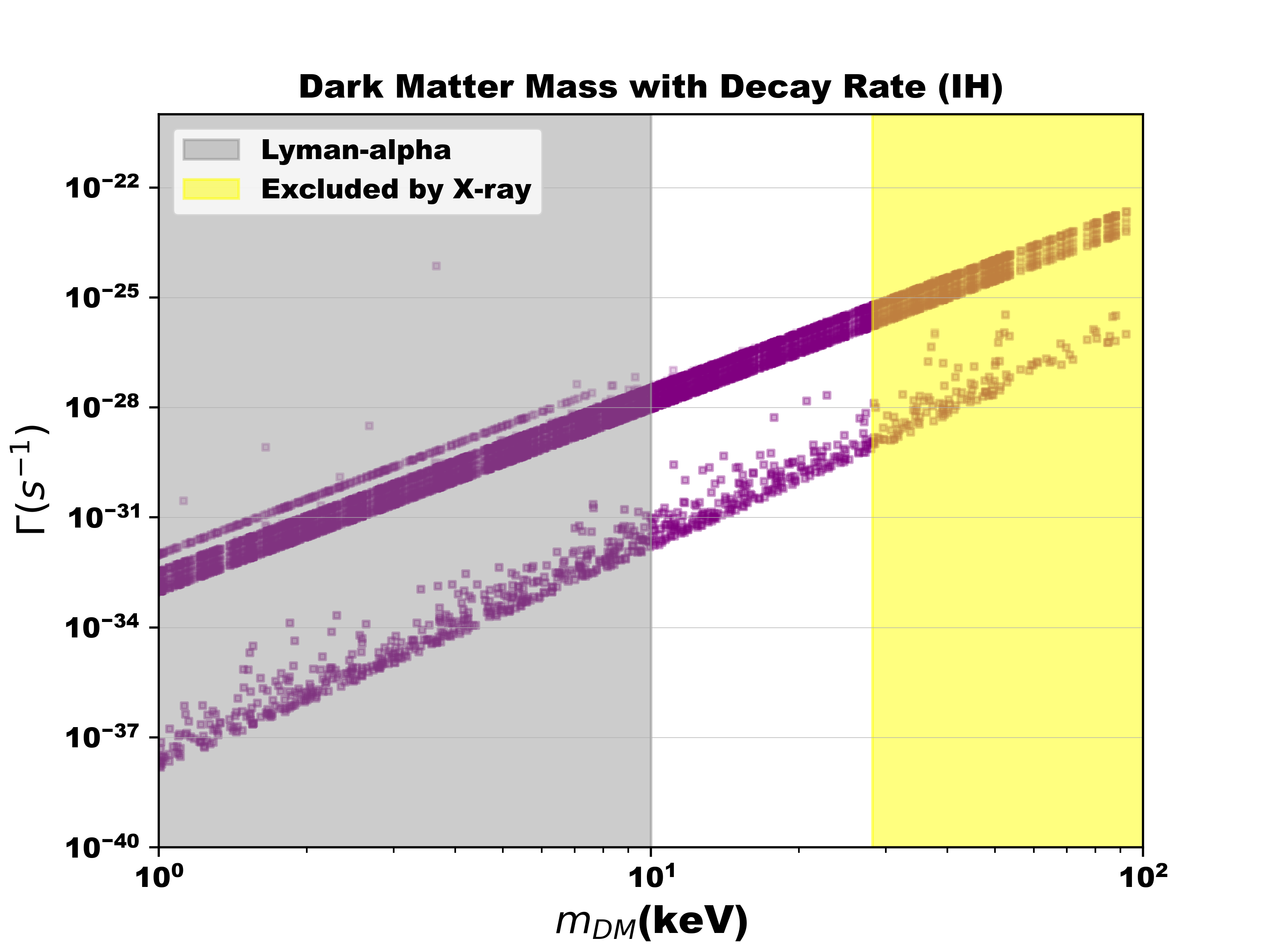}
		\caption{Variation of dark matter mass with decay rate of the dark matter candidate for weight $k_{Y}=10$.}
		\label{f103}
	\end{figure}
	\section{Numerical Analysis and Results}\label{w76}
\begin{figure}[h]
	\centering
	\includegraphics[scale=0.45]{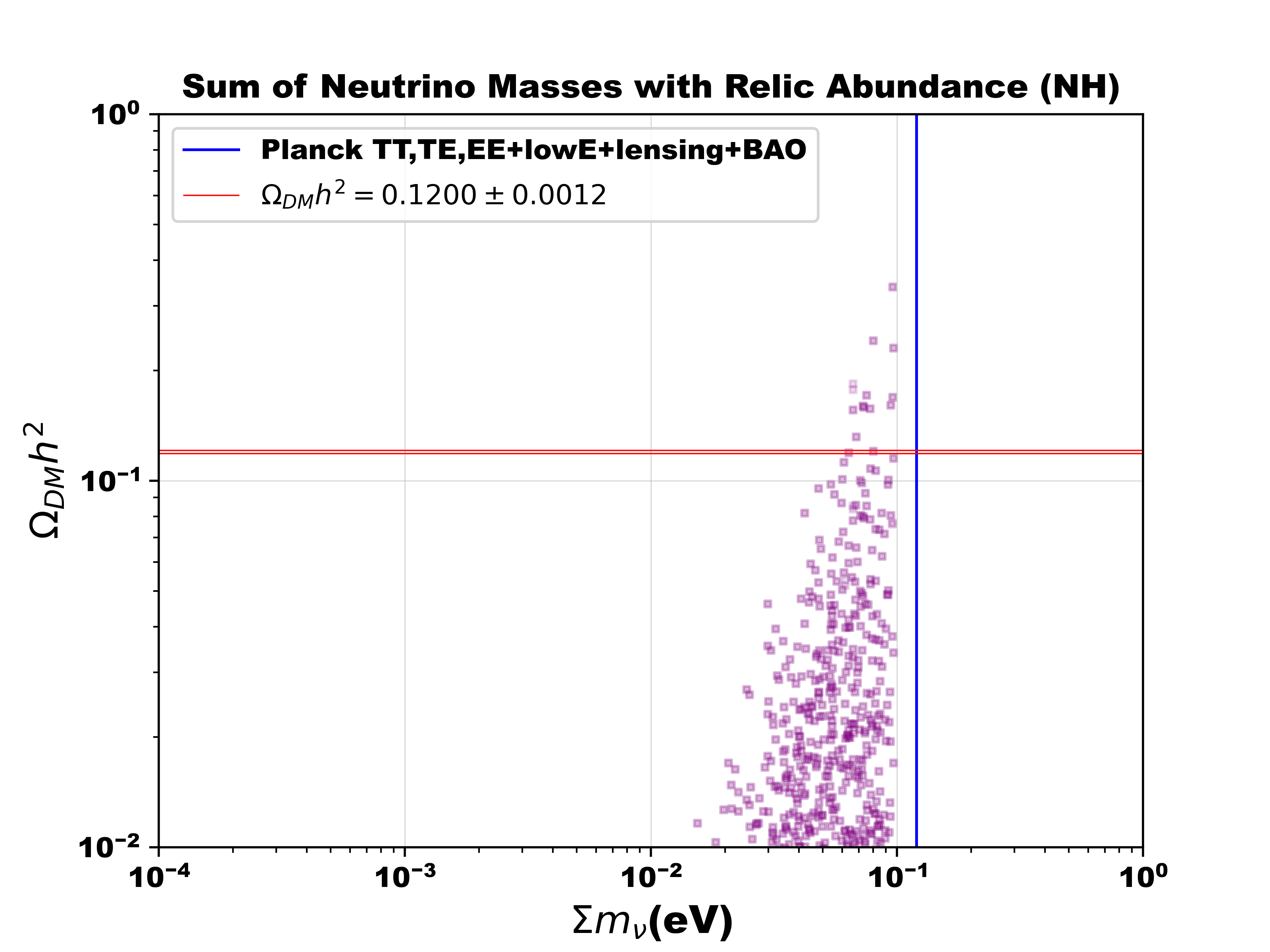}
	\includegraphics[scale=0.45]{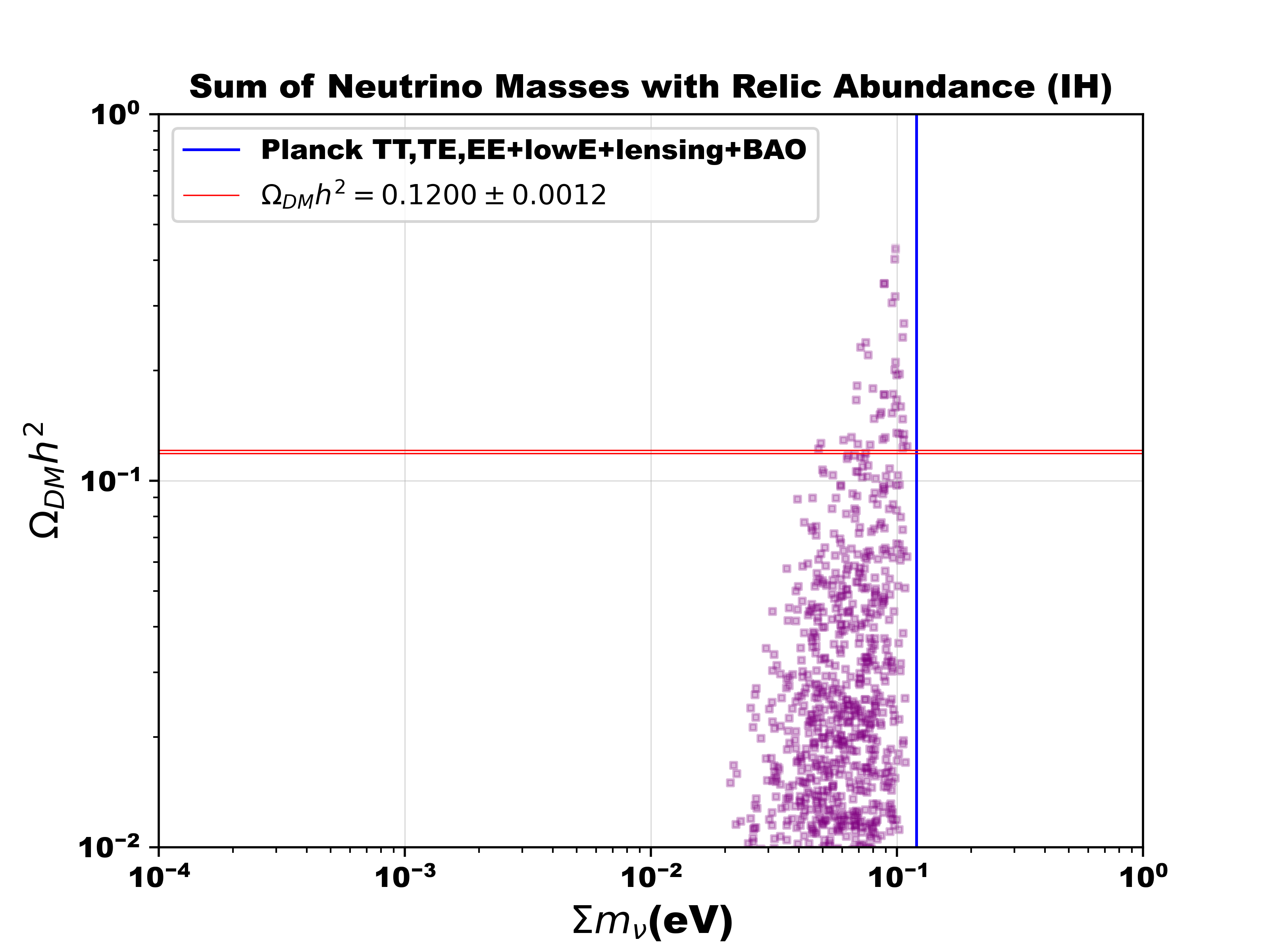}
	\caption{The relic abundance as a function of sum of neutrino masses and the red band shows the predicted value of relic abundance, $\Omega_{DM}h^{2}=0.1200\pm 0.0012$, the blue line corresponds to the Planck limit on sum of neutrino masses for $k_{Y}=4$.}
	\label{f44}
\end{figure}

\begin{figure}[h]
	\centering
	\includegraphics[scale=0.45]{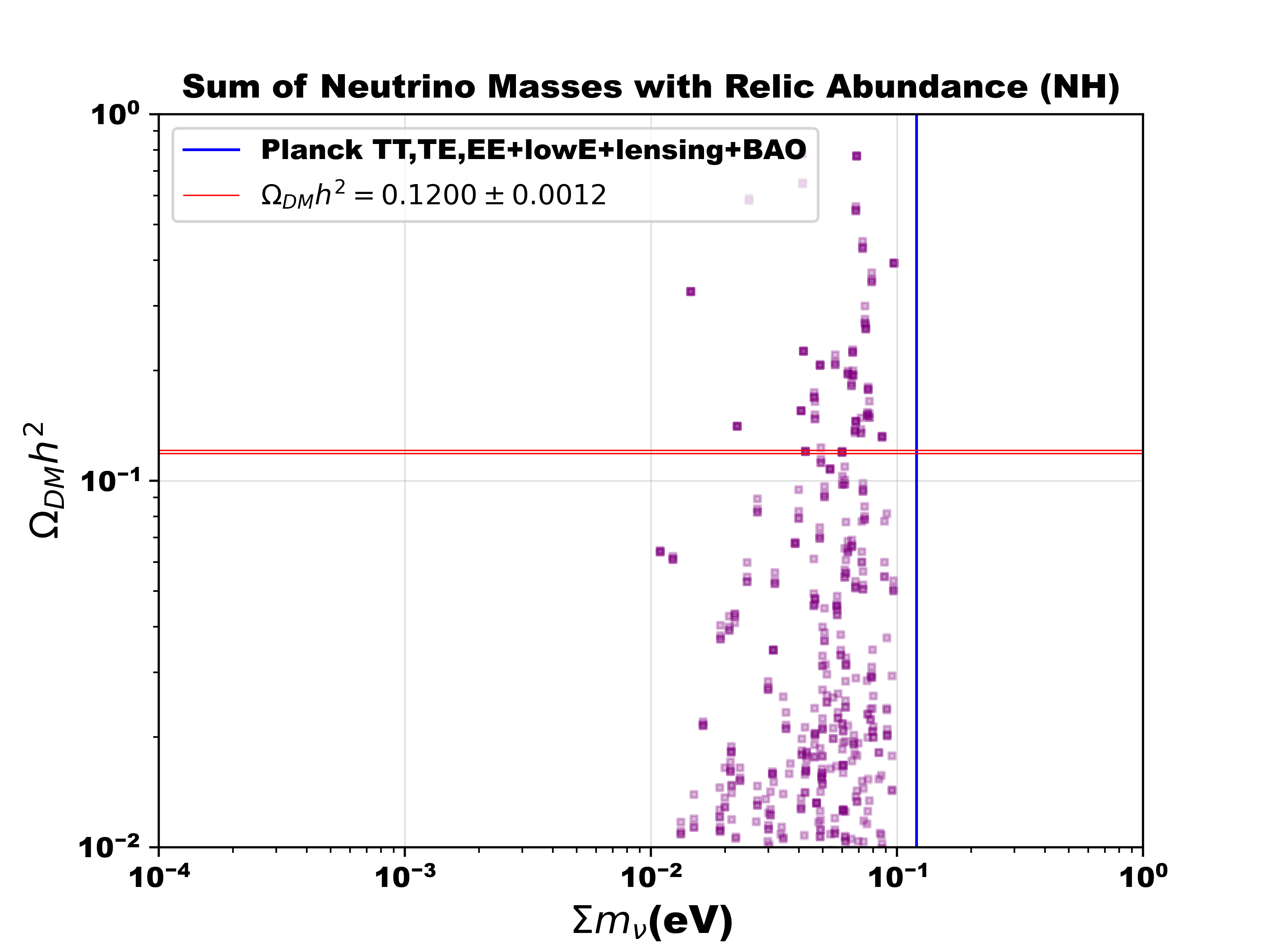}
	\includegraphics[scale=0.45]{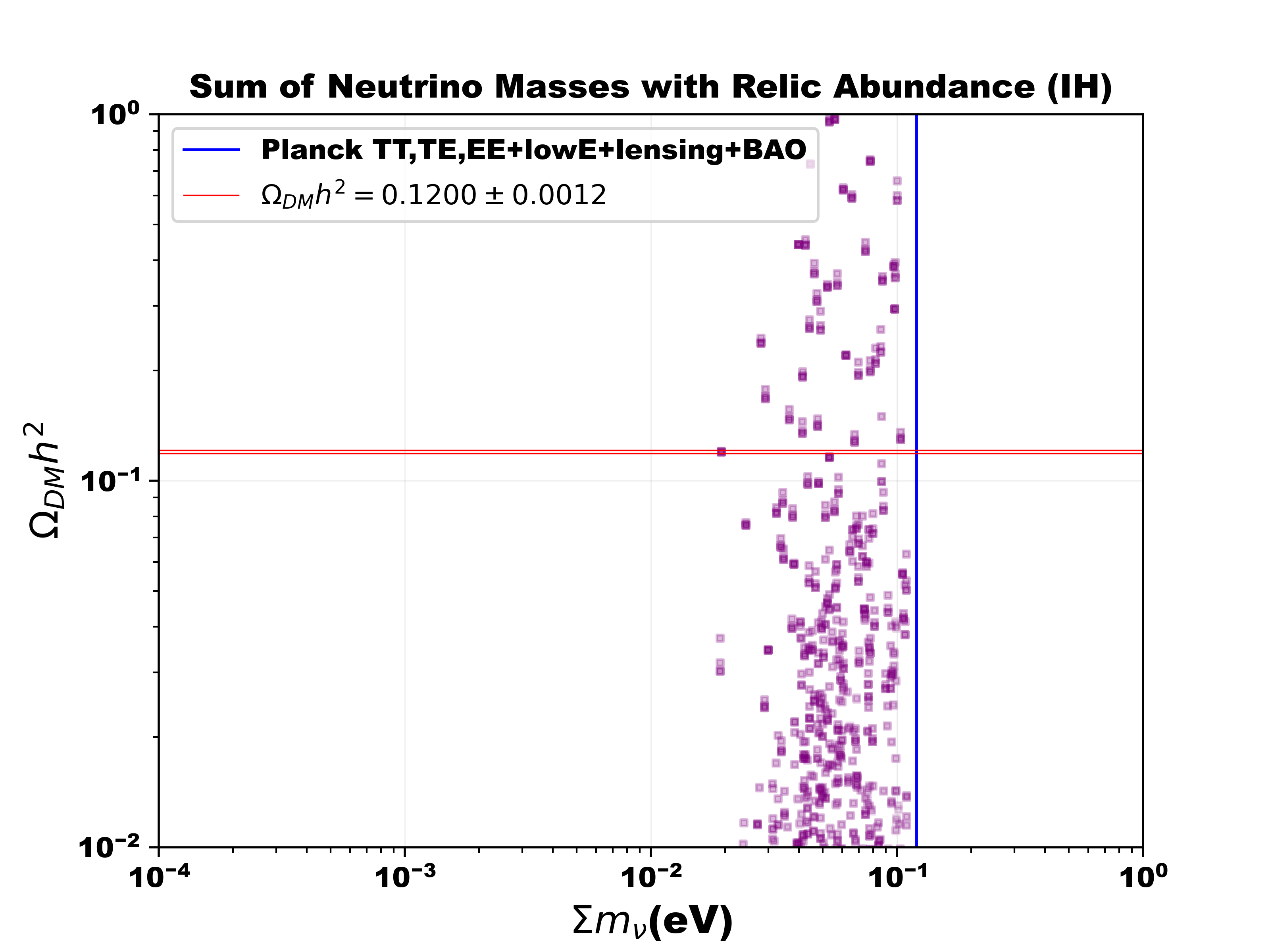}
	\caption{The relic abundance as a function of sum of neutrino masses and the red band shows the predicted value of relic abundance, $\Omega_{DM}h^{2}=0.1200\pm 0.0012$, the blue line corresponds to the Planck limit on sum of neutrino masses for $k_{Y}=8$.}
	\label{f84}
\end{figure}

\begin{figure}[h]
	\centering
	\includegraphics[scale=0.45]{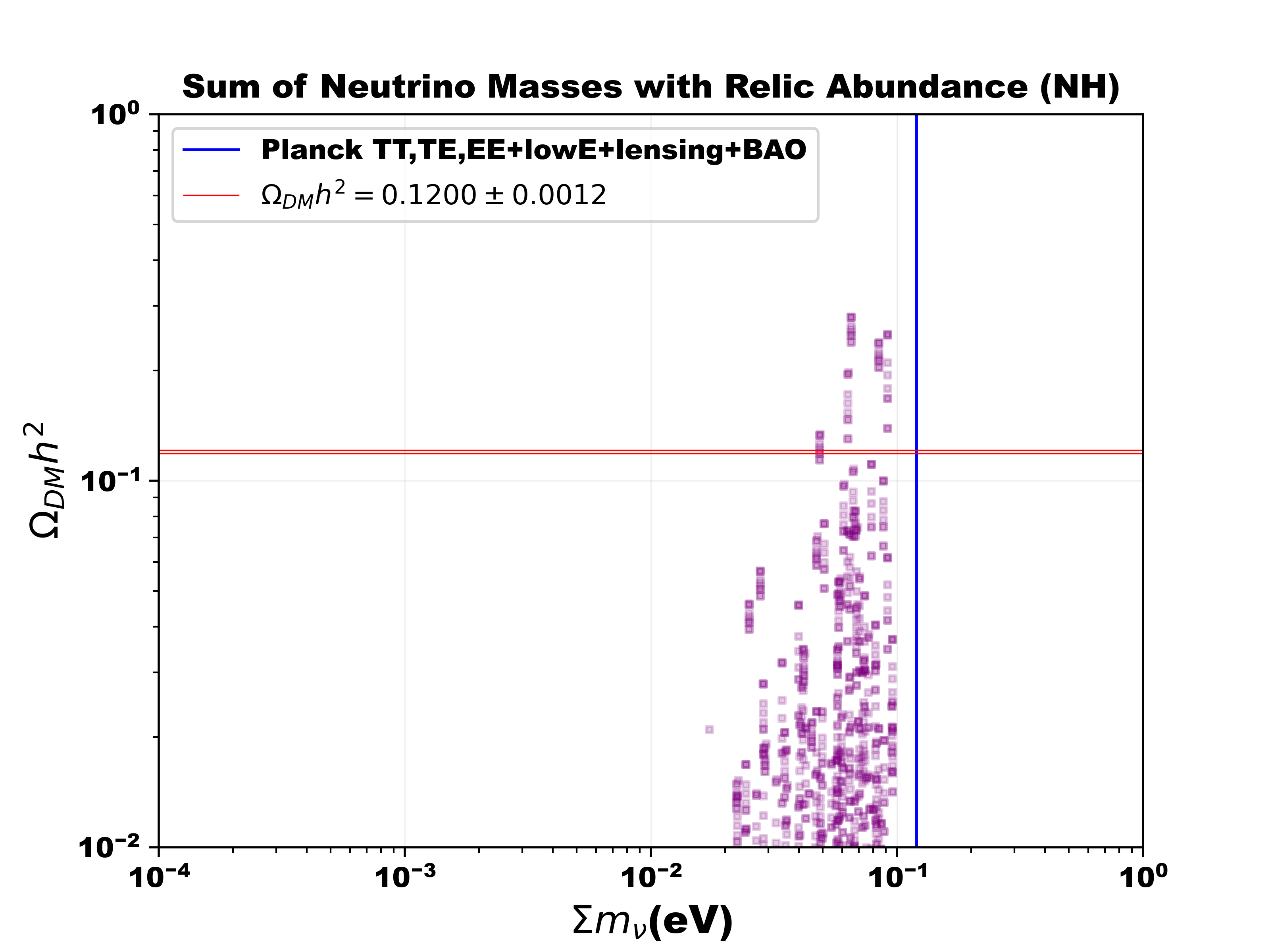}
	\includegraphics[scale=0.45]{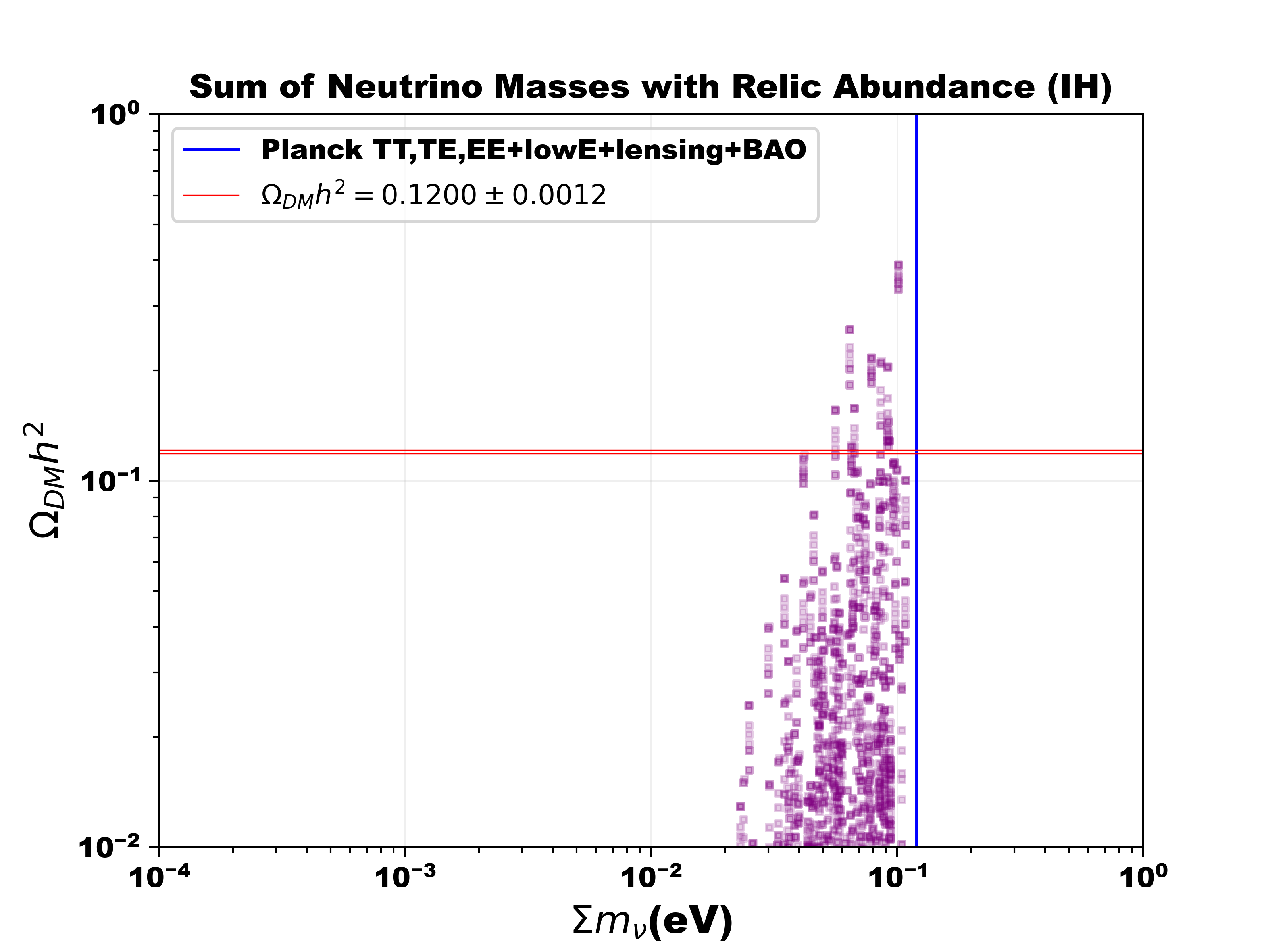}
	\caption{The relic abundance as a function of sum of neutrino masses and the red band shows the predicted value of relic abundance, $\Omega_{DM}h^{2}=0.1200\pm 0.0012$, the blue line corresponds to the Planck limit on sum of neutrino masses for $k_{Y}=10$.}
	\label{f104}
\end{figure}

To begin with, first for each of the weights, charge and modular weights for each of the particles within the model have been assigned. While assigning the modular weights, the criteria that sum of modular weights in each terms of the Lagrangian should be zero. After determining the respective mass matrices, the active-sterile mixing angle, $sin^{2}2\theta_{DM}$ has been calculated. For calculation of relic abundance and the decay rate of dark matter candidate, equations \eqref{7e12} to \eqref{7e15} have been used. The results obtained have been depicted using figures \ref{f41} to \ref{f103}. However, not only the associated parameters, but the variation of sum of neutrino masses have also been plotted against the relic abundace of dark matter and the results for all the weights have been shown in figures \ref{f44} to \ref{f104}.
As stated in section \ref{www73}, the decay of the right handed neutrino, $N_{1}$ is considered responsible for the production of required CP asymmetry and which is determined using equation \eqref{7e18} and finally the baryon asymmetry parameter $n_{B}$ is calculated using the equation \eqref{7e19}. 
\begin{equation}
	\label{7e18}
	\epsilon_{1} \equiv \frac{3. M_{N_{1}}}{16\pi (M_{D}^{\dagger}M_{D})_{11}}\Bigg[\frac{Im[(M_{D}^{\dagger}M_{D})^{2}_{21}]}{M_{N_{2}}}+\frac{Im[(M_{D}^{\dagger}M_{D})^{2}_{31}]}{M_{N_{3}}}\Bigg]
\end{equation}
As such, to calculate the CP-asymmetry parameter, we first need to determine the structure of $M_{D}$ and masses of the RH neutrinos. The net lepton asymmetry is finally given by the baryon asymmetry parameter $\eta$ given by,
\begin{equation}
	\label{7e19}
	\eta_B \approx -0.96 \times 10^{-2}\sum_{i}(k_{i}\epsilon_{i})
\end{equation}
$k_i$ being the efficiency factors measuring the washout effects. Some parameters are needed to defined as,
\begin{equation}
	\label{7e20}
	K_i \equiv \frac{\Gamma_{i}}{H}
\end{equation}
Equation \eqref{7e20} is defined at temperature $T=M_{i}$. The Hubble's constant is given by, $H\equiv \frac{1.66\sqrt{g_{*}}T^{2}}{M_{Planck}}$, where, $g_{*}=107$ and $M_{Planck} = 1.2 \times 10^{19}$ GeV is the Planck mass. Decay width is estimated using \eqref{g1}. The efficiency factors $k_i$ can be calculated using the formula \cite{Blanchet:2008pw},
\begin{itemize}
	\item  For $K\geq 10^{6}$,
	\begin{equation*}
		k_{i} \approx \sqrt{0.1K}exp\Bigg(\frac{4}{3(0.1K)^{0.25}}\Bigg)
	\end{equation*} 
	\item  For $10 \leq K\leq 10^{6}$,
	\begin{equation*}
		k_{i} \approx \frac{0.3}{K ln(K)^{0.6}}
	\end{equation*}
	\item For $0 \leq K\leq 10$
	\begin{equation*}
		k_{i} \approx \frac{1}{2\sqrt{K^{2}+9}}
	\end{equation*}
\end{itemize}
\begin{figure}[h]
	\centering
	\includegraphics[scale=0.45]{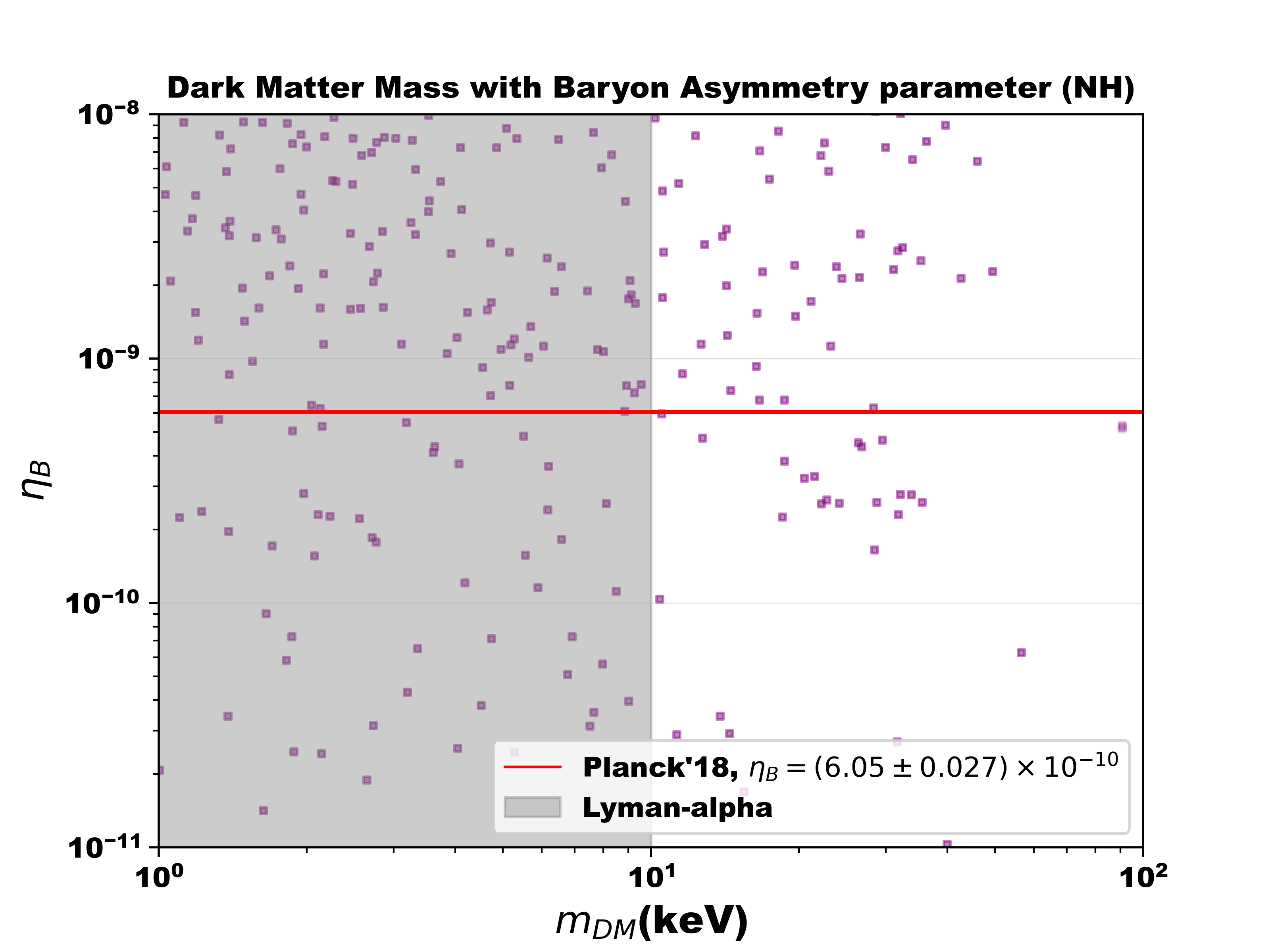}
	\includegraphics[scale=0.45]{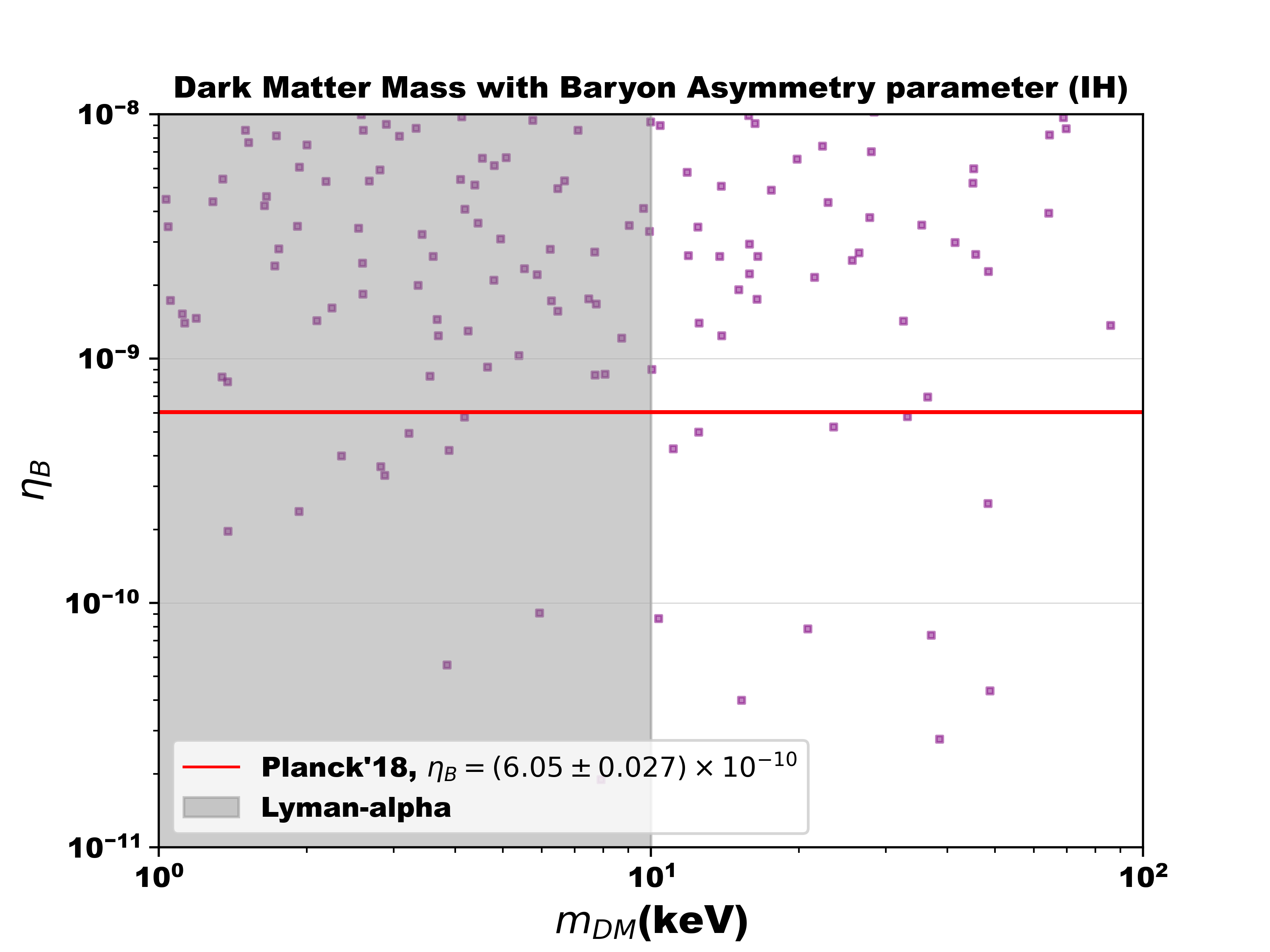}
	\caption{Baryon asymmetry parameter as a function of dark matter mass where the red band depicts the baryon asymmetry parameter denoted by $\eta_B=(6.05\pm 0.027)\times 10^{-10}$ for weight $k_{Y}=4$.}
	\label{f45}
\end{figure}
\begin{figure}[h]
	\centering
	\includegraphics[scale=0.45]{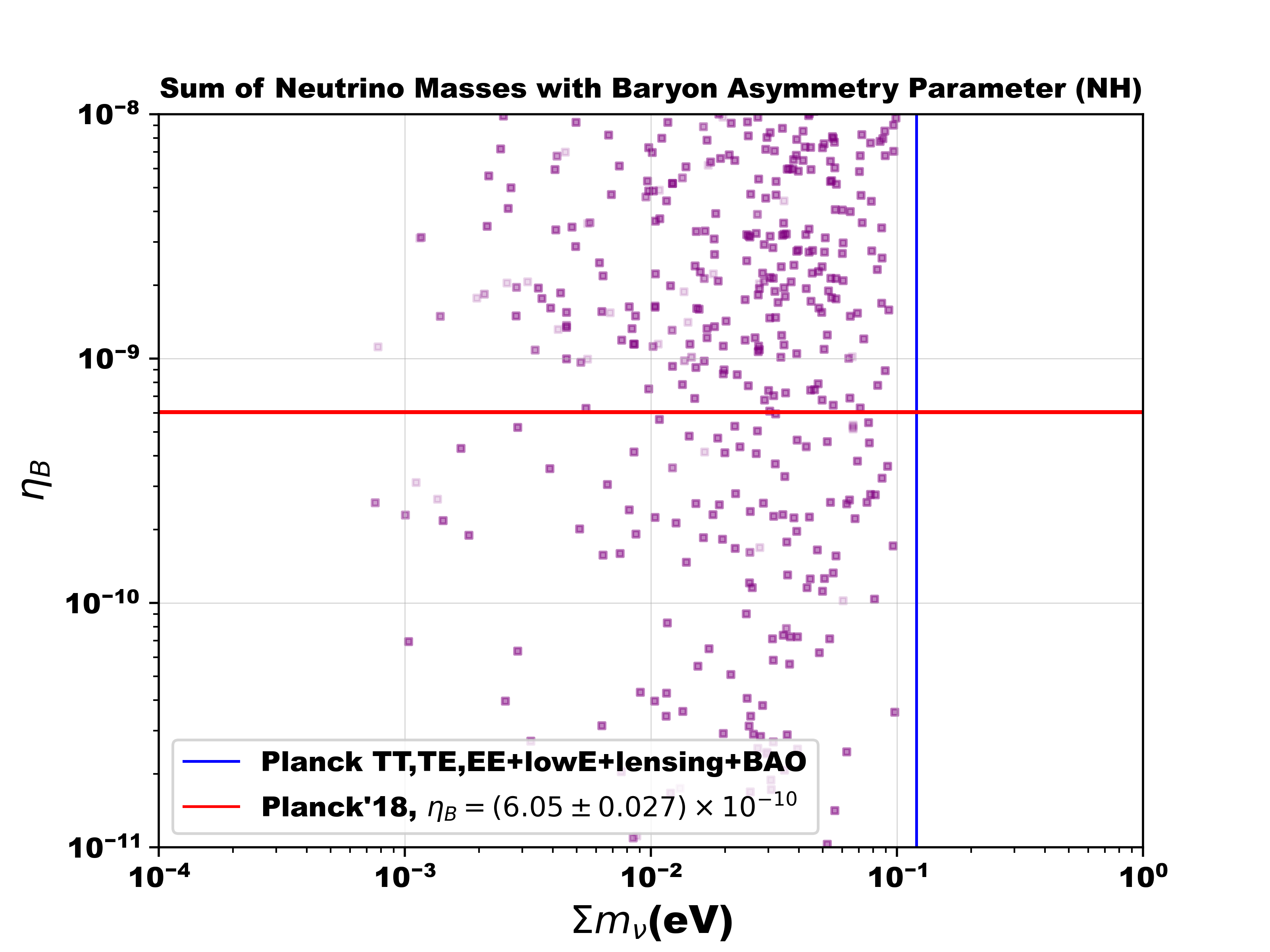}
	\includegraphics[scale=0.45]{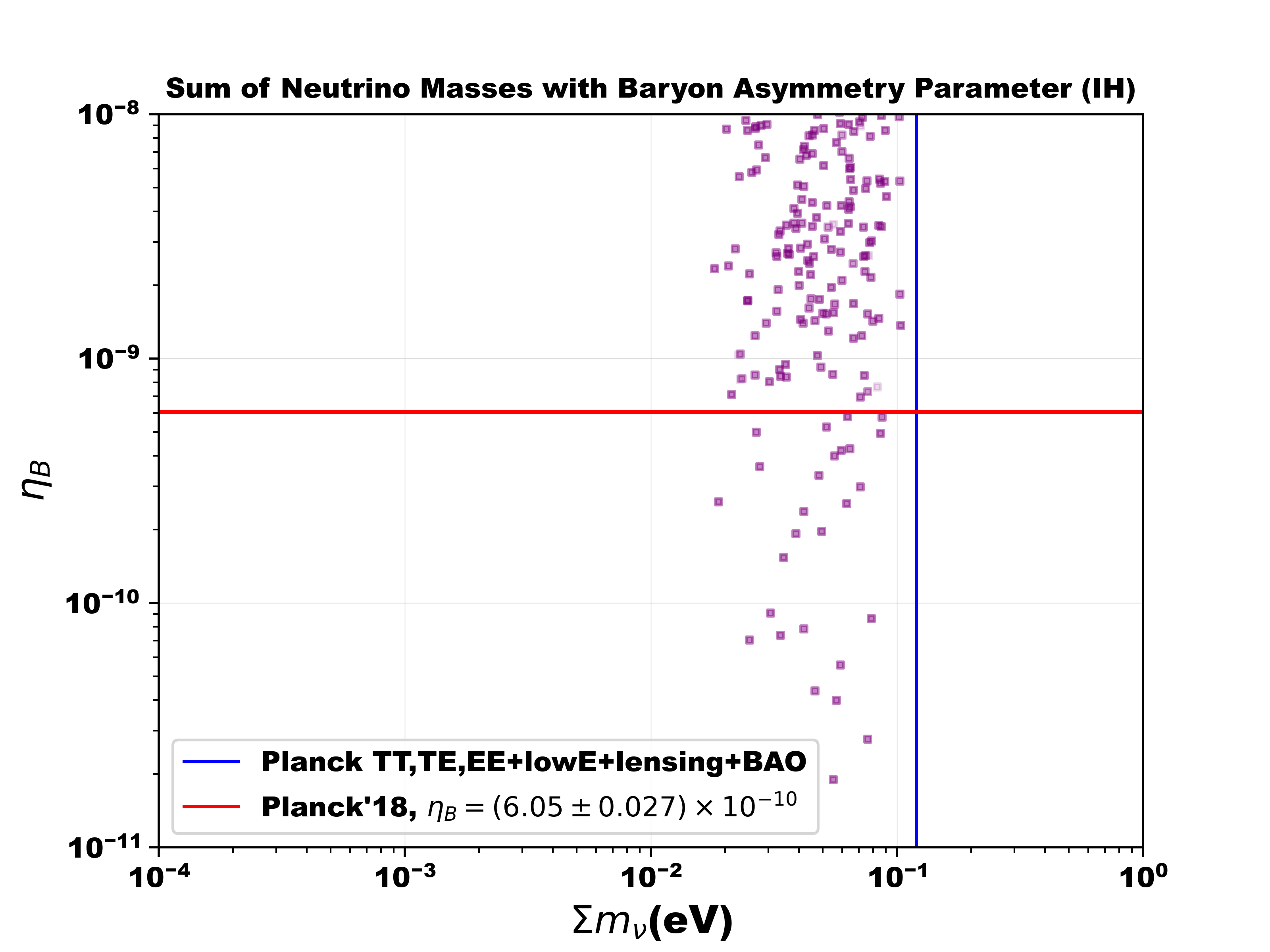}
	\caption{Baryon asymmetry parameter as a function of dark matter mass where the red band depicts the baryon asymmetry parameter denoted by $\eta_B=(6.05\pm 0.027)\times 10^{-10}$ for weight $k_{Y}=4$.}
	\label{f46}
\end{figure}
	\begin{figure}[h]
		\centering
		\includegraphics[scale=0.45]{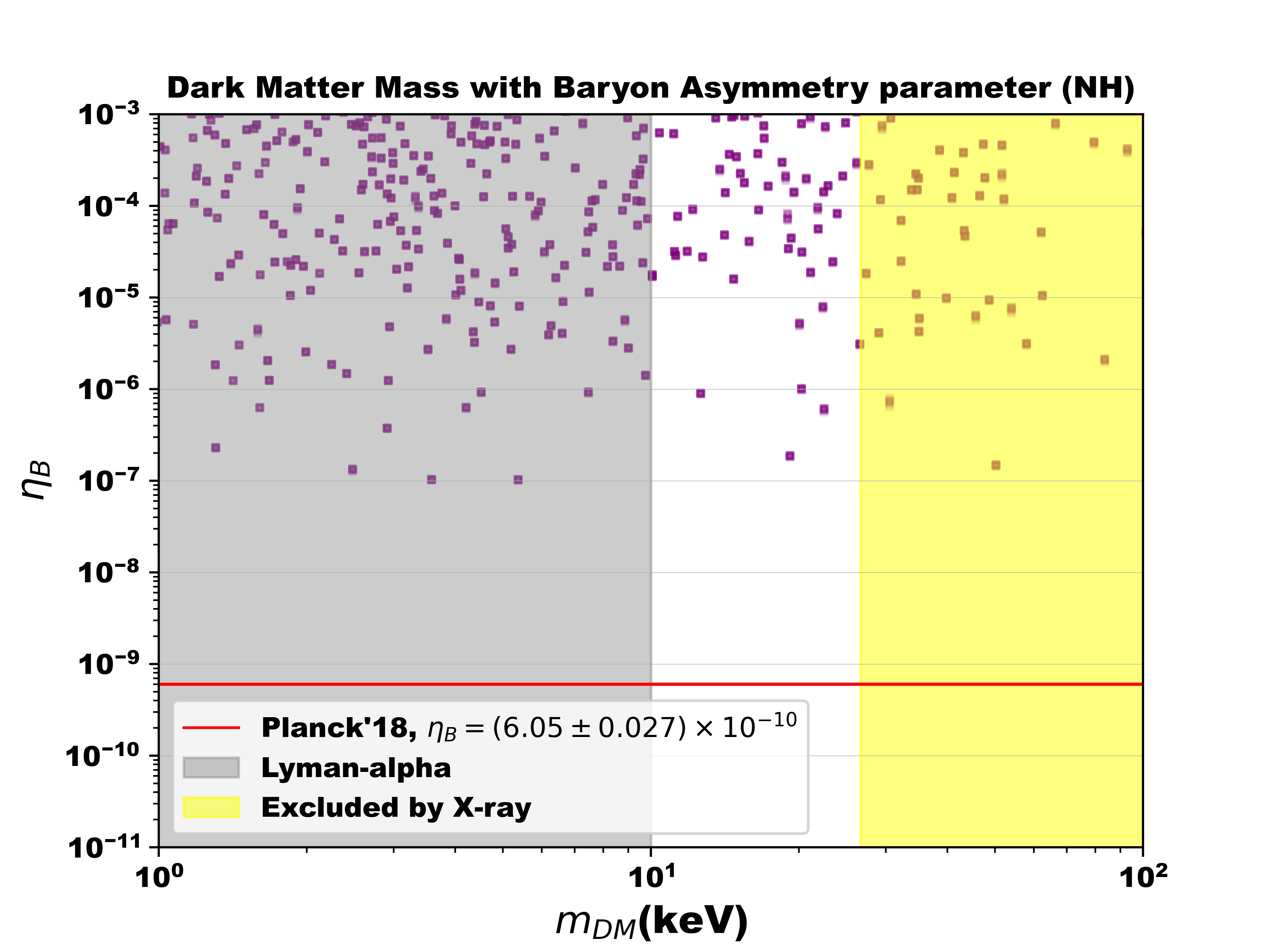}
		\includegraphics[scale=0.45]{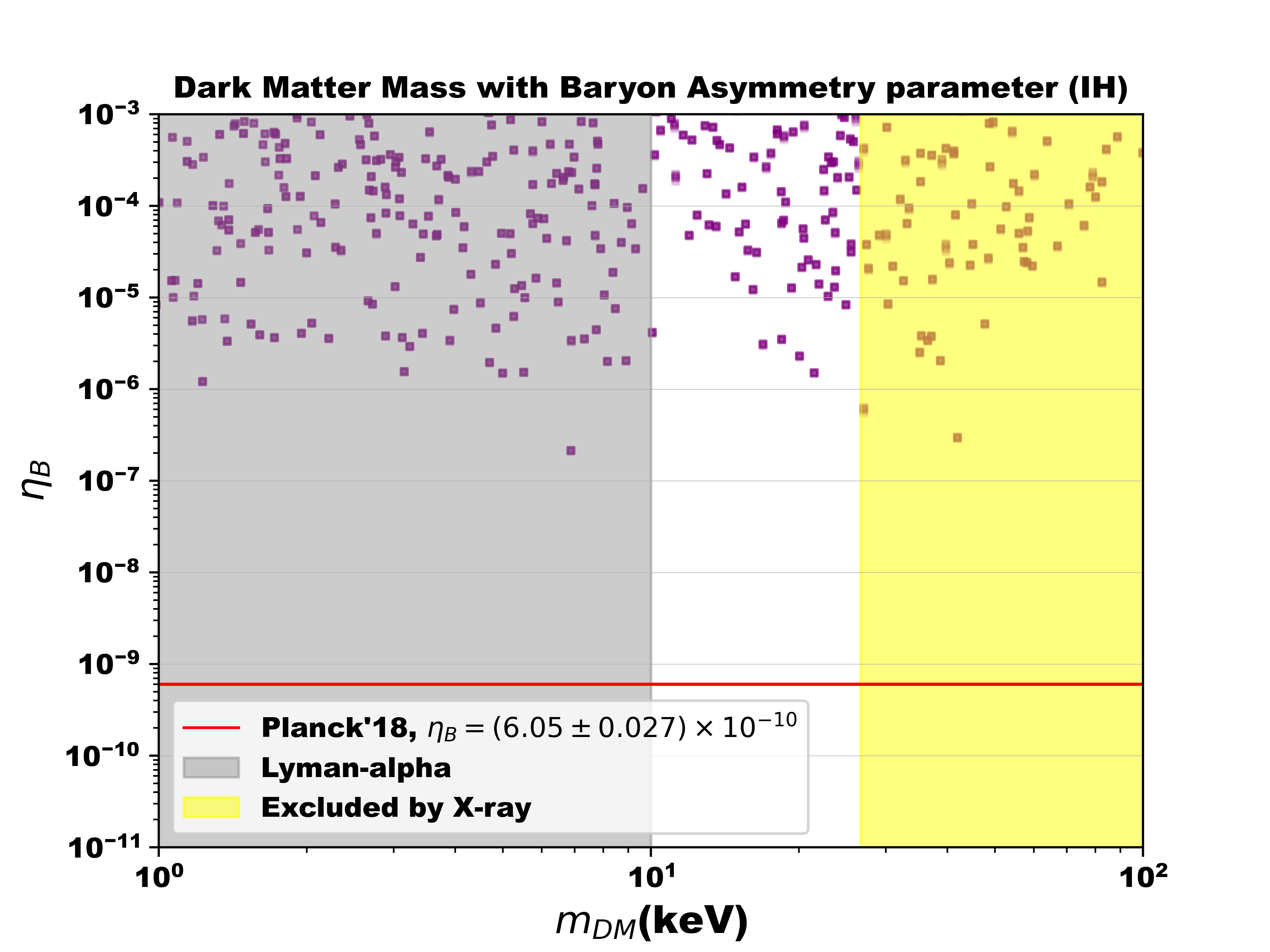}
		\caption{Baryon asymmetry parameter as a function of dark matter mass where the red band depicts the baryon asymmetry parameter denoted by $\eta_B=(6.05\pm 0.027)\times 10^{-10}$ for weight $k_{Y}=8$.}
		\label{f85}
	\end{figure}
	\begin{figure}[h]
		\centering
		\includegraphics[scale=0.45]{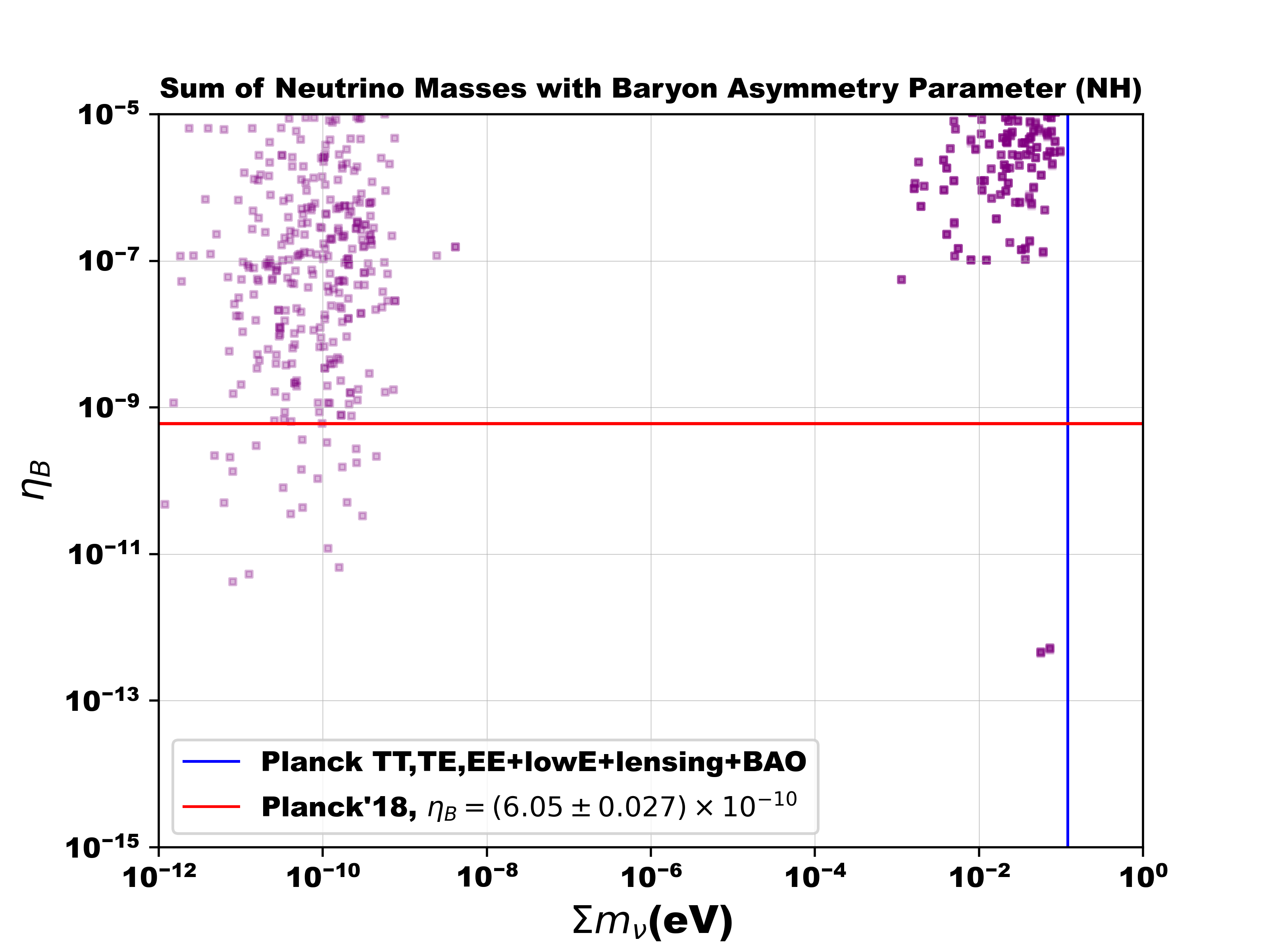}
		\includegraphics[scale=0.45]{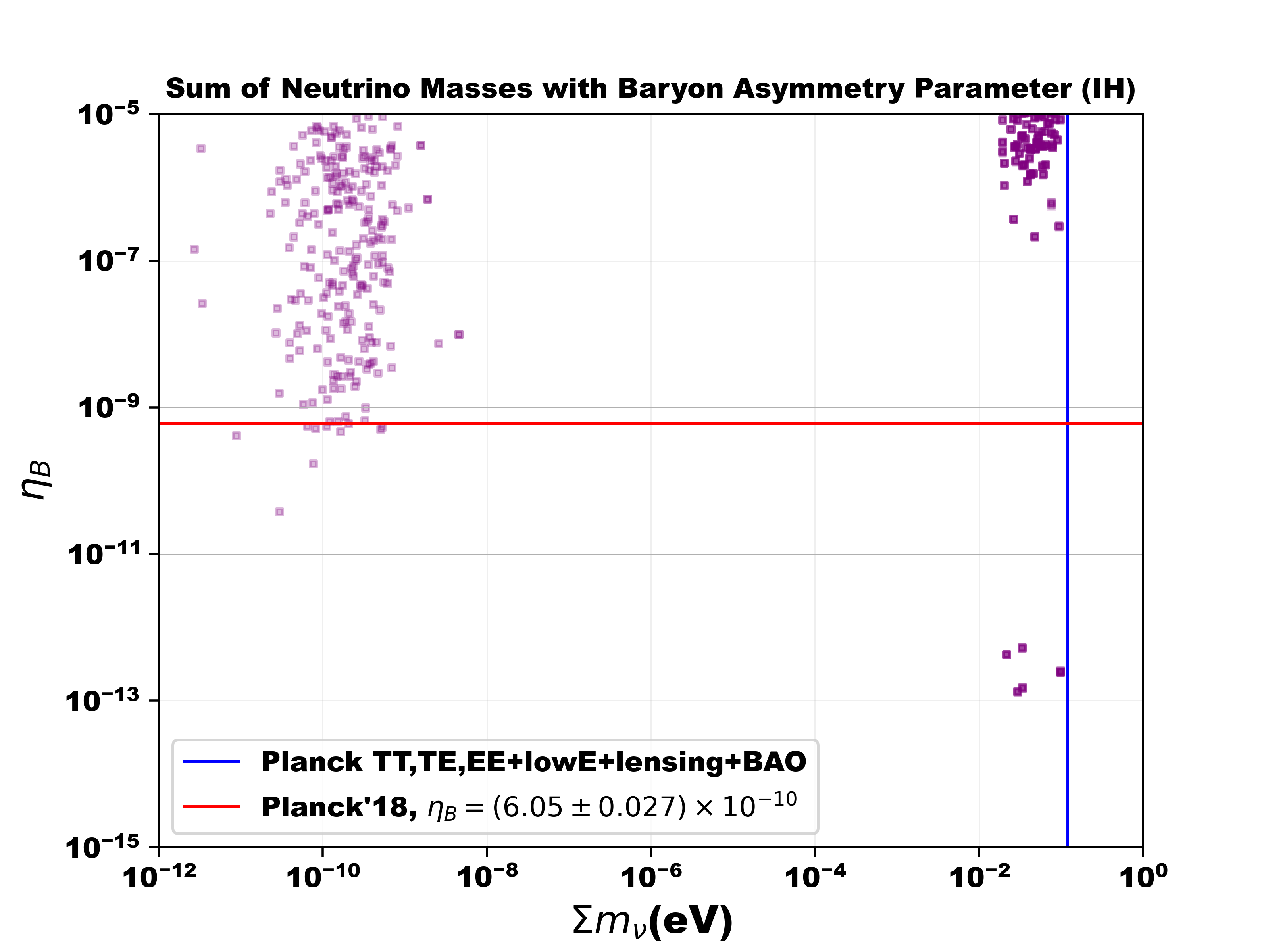}
		\caption{Baryon asymmetry parameter as a function of dark matter mass where the red band depicts the baryon asymmetry parameter denoted by $\eta_B=(6.05\pm 0.027)\times 10^{-10}$ for weight $k_{Y}=8$.}
		\label{f86}
\end{figure}
	\begin{figure}[h]
	\centering
	\includegraphics[scale=0.45]{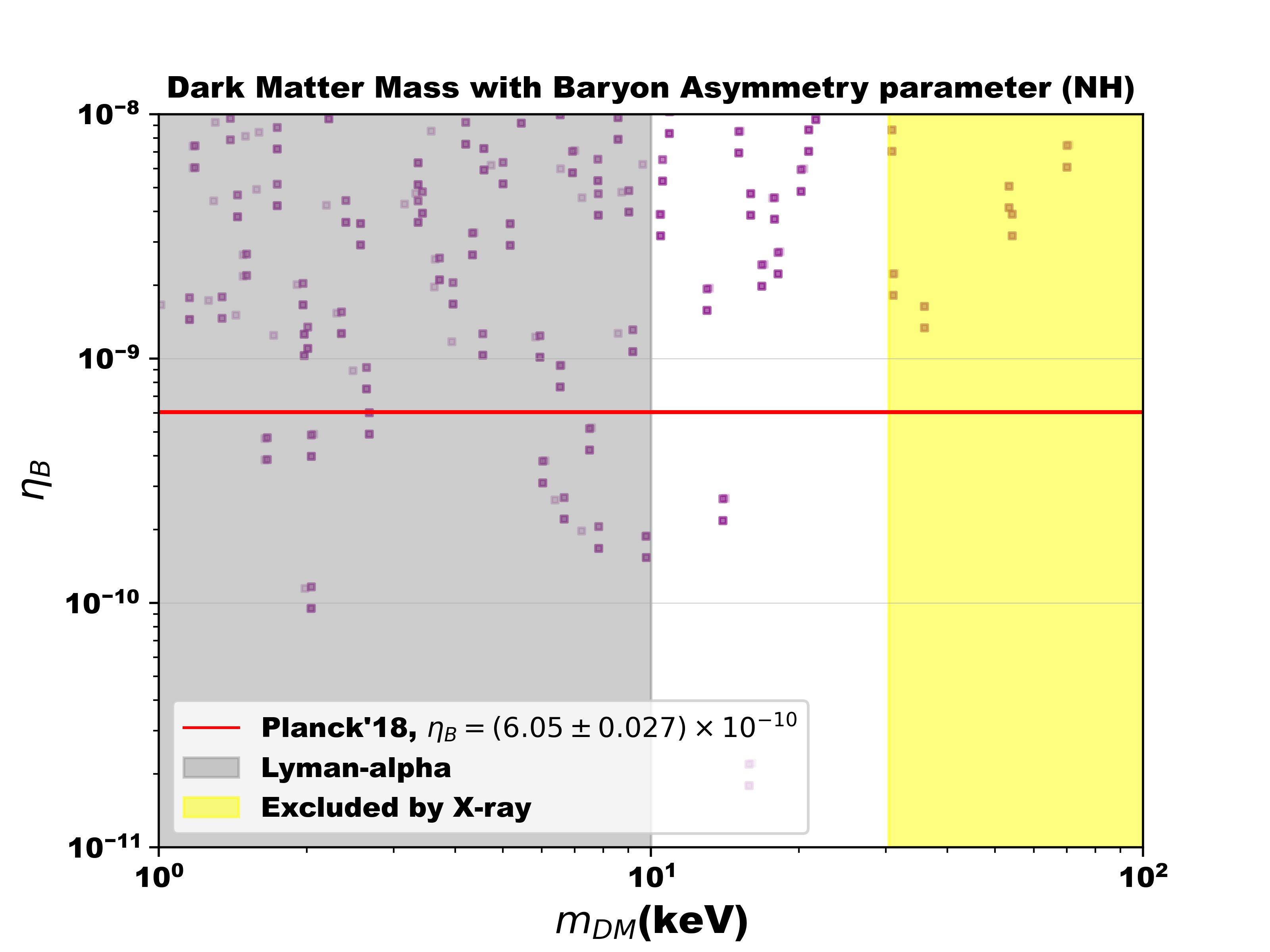}
	\includegraphics[scale=0.45]{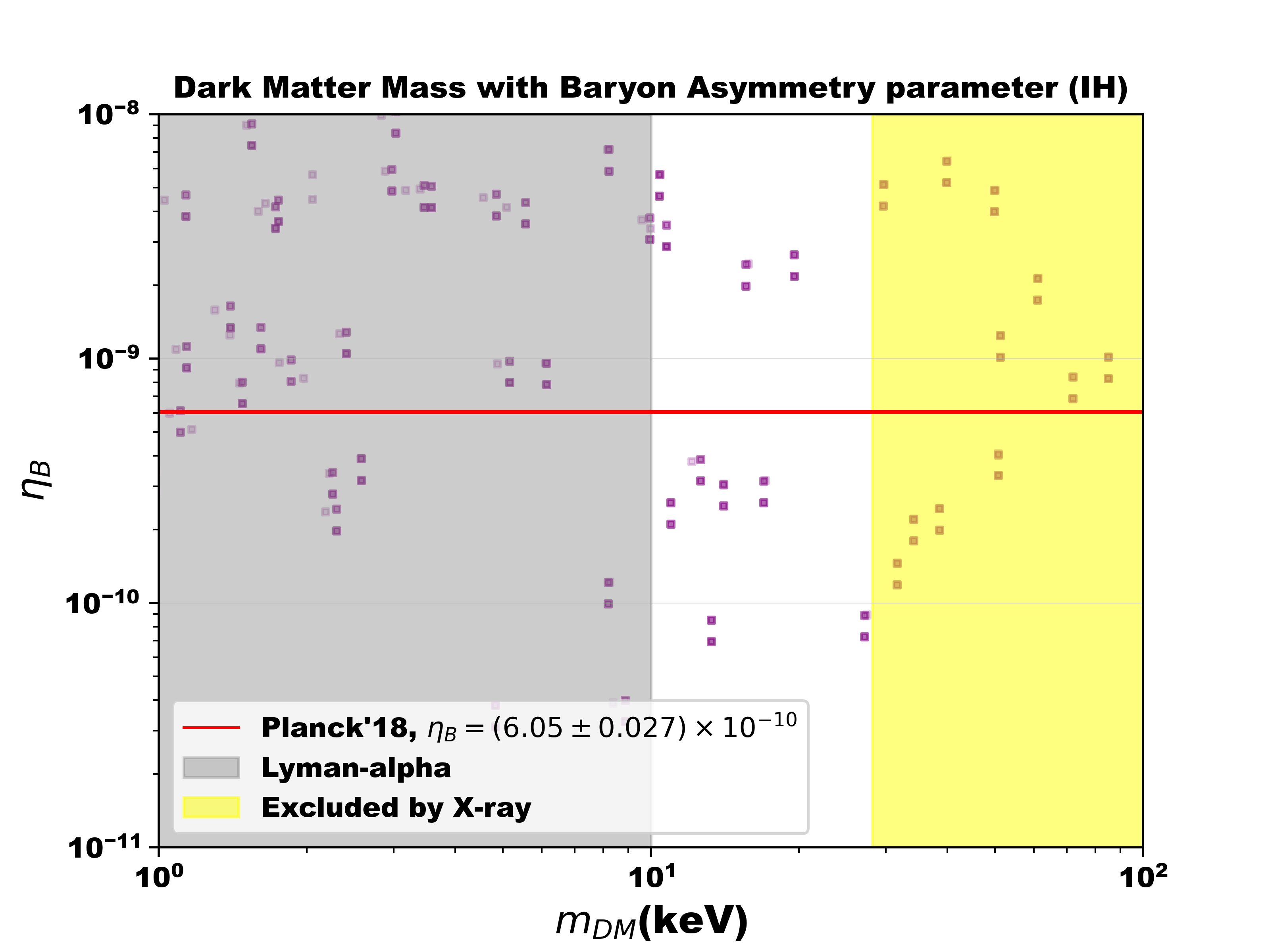}
	\caption{Baryon asymmetry parameter as a function of dark matter mass where the red band depicts the baryon asymmetry parameter denoted by $\eta_B=(6.05\pm 0.027)\times 10^{-10}$ for weight $k_{Y}=10$.}
	\label{f105}
\end{figure}
\begin{figure}[h]
	\centering
	\includegraphics[scale=0.45]{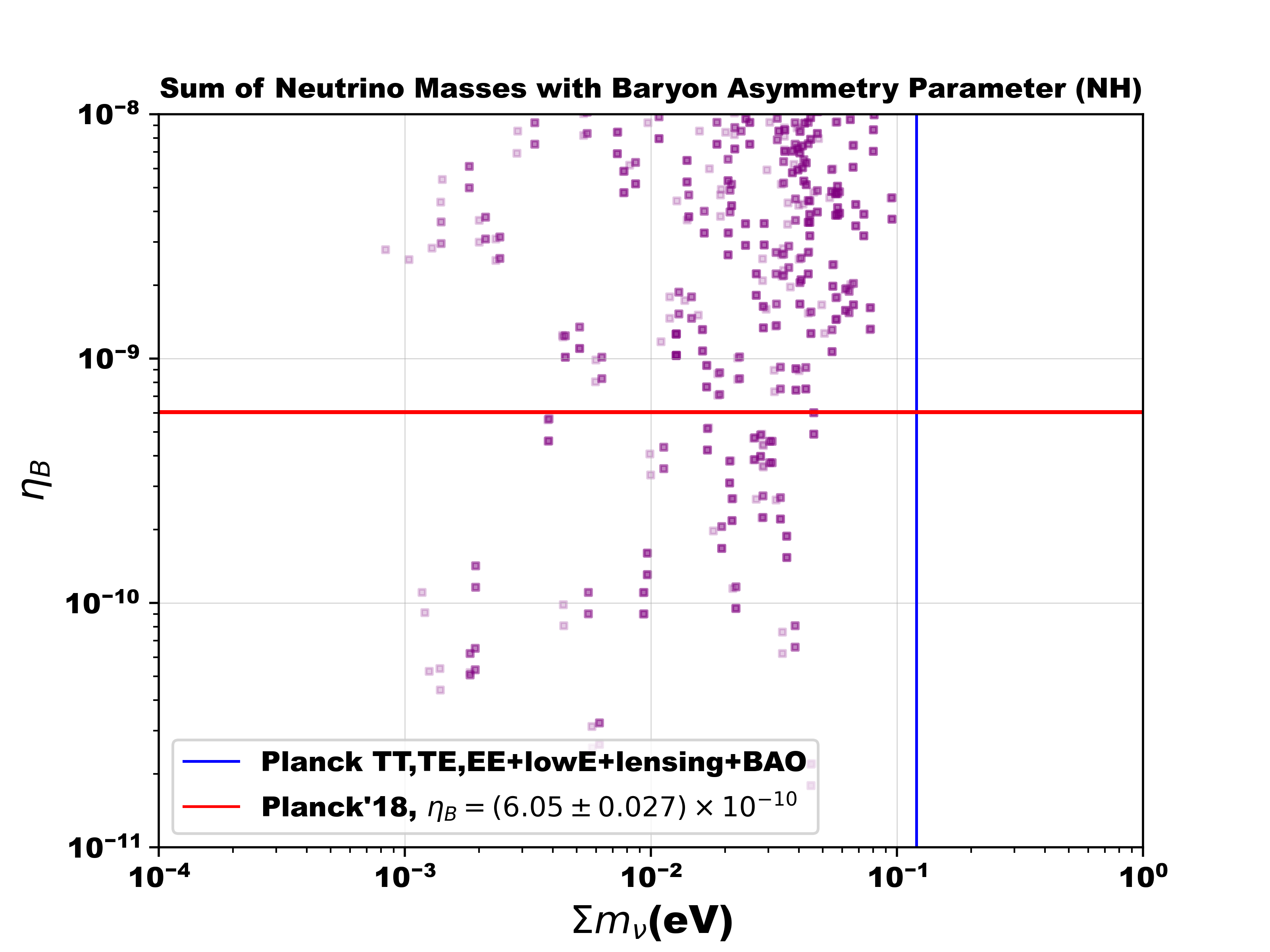}
	\includegraphics[scale=0.45]{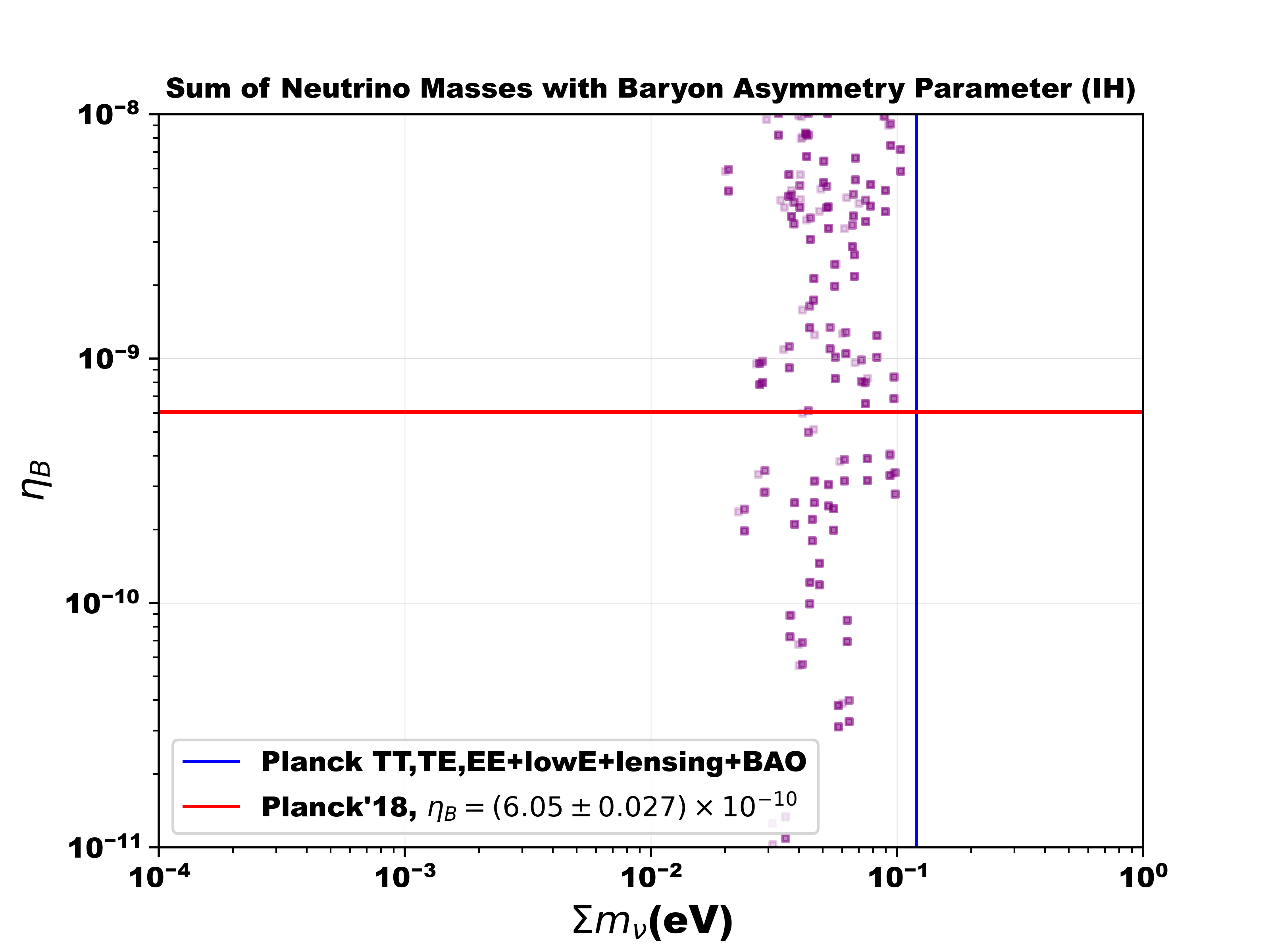}
	\caption{Baryon asymmetry parameter as a function of dark matter mass where the red band depicts the baryon asymmetry parameter denoted by $\eta_B=(6.05\pm 0.027)\times 10^{-10}$ for weight $k_{Y}=10$.}
	\label{f106}
\end{figure}
\section{Discussion and Conclusion}\label{w77}
The present work aims to investigate the extension of doublet LRSM with a sterile neutrino per generation so as to take into account the study of sterile neutrino as a dark matter candidate. However, implication of modular symmetry has been cariied out for different weights of the modular form $Y$. What has been observed is that although the resulting neutrino mass matrices are of the same texture for different weights, the results obtained are much different which might be due to the different elements in each of the mass matrices. The Yukawa couplings are represented as modular forms expressed as expansions of $q$. The values of the Yukawa couplings $(Y_1,Y_2,Y_3)$ are calculated using the relation $M_{\nu}=U_{PMNS}m_{\nu}^{diag}U_{PMNS}^{T}$. After determination of the modular forms, the respective mass matrices which are expressed in terms of the modular forms $Y_{1},Y_{2},Y_{3}$ are determined and hence finally, the respective phenomenological parameters are calculated using the formulae stated throughout the manuscript.\\
The following points thoroughly discusses the results and observations obtained in the current work.
\begin{itemize}
	\item Left-Right Symmetric Model has been extended with a sterile neutrino per generation and the scalar sector consists of a Higgs bidoublet $\phi(1,2,2,0)$ and two scalar doublets $H_{L}(1,2,1,-1)$ and $H_{R}(1,1,2,-1)$. The resulting neutrino mass in the current scenario is given by the double seesaw mechanism. The question is why double seesaw. It directly points to the natural origin of small masses for sterile neutrinos at high $SU(2)_{R}$ breaking scales.
	\item Generally in studies where LRSM is extended with sterile neutrinos for the study of dark matter, the lightest of the same is considered to be a suitable dark matter candidate. However, in the current scenario all the three sterile neutrinos possess the same mass, and as such the concept of multi-component dark matter comes into action, where all the three sterile neutrinos can take part as a suitable dark matter candidate.
	\item After constructing the superpotential for the model, charge and modular weights corresponding to each particle has been assigned, keeping in mind the invariance of the superpotential under the model gauge group. 
	\item To discuss about the results obtained in the study, let's begin with the observation of the respective plots showing variations of different phenomenological parameters. Figure \ref{f41} shows the variation of active-sterile mixing as a function of mass of the dark matter candidate for $k_{Y}=4$, and it has been observed that the data points do not satisfy the constraints imposed by Lyman-$\alpha$ forest and X-ray constraints simultaneously. For both normal and inverted hierarchies, it has been observed that the mixing angle falls within the range of around $10^{-11}$ to less than $10^{-10}$. In some studies, it has been mentioned that for a good dark matter candidate the mixing angle must range from around $10^{-12}$ to $10^{-8}$, and as such the mixing angle is successfully satisfying the criteria, but as a matter of fact, for the mentioned weight a set of values for allowed mass range of the dark matter candidate has not been observed.
	\item Figures \ref{f42} and \ref{f43} shows the variation of dark matter mass with relic abundance and decay rate respectively for weight $k_{Y}=4$. Although the gray region depicts the Lyman-$\alpha$ constraints, but a particular mass range satisfying also the X-ray constraints was not observed and as such, the plot does not predict a specific area of interest. However, it has been observed that there are data points satisfying the predicted value of relic abundance of dark matter. The decay rate ranges from around $10^{-33}$ to $10^{-22}$ which is almost negligible and is a satisfactory result.
	\item For weight $k_{Y}=6$ although the resulting neutrino mass matrix has been derived, but not suitable results were predicted for the same and as such, there are not plots for the case of $k_{Y}=6$.
	\item From figure \ref{f81}, it is evident that the allowed range for dark matter mass falls in between $10keV$ to $33.98keV$ for normal hierarchy, and from $10keV$ to $26.632keV$ for inverted hierarchy. The data points falling within this range satisfies both the Lyman-$\alpha$ and X-ray constraints. Similarly figures \ref{f82} and \ref{f83} shows relic abundance and decay rate as a function of dark matter mass and for relic abundance the observed value of $\Omega_{DM}h^{2}$ is found to be satisfied within the allowed mass range.
	\item For weight $k_{Y}=10$, the mixing angle is found to lie from $10^{-16}$ to around $10^{-13}$ for both normal and inverted hierarchy as depicted in figure \ref{f101}. It is comparatively smaller as compared to weights 4 and 8, however, an allowed parameter space has been obtained for mass of dark matter satisfying both the constraints. For normal hierarchy, the mass falls upto $30.39keV$ and for inverted hierarchy, it becomes $28.1577keV$. From figure \ref{f102} it is evident that the data points satisfying the value for relic abundance falls in the mass region falling in the X-ray constraints and hence, the result is not satisfactory. Figure \ref{f103} shows the variation of dark matter mass with the decay rate.
	\item Coming to the variation of relic abundance as a function of sum of neutrino masses, for different weights it has been found that the data points pertaining to observed value of relic abundance also satisfy the Planck limit(lensing + BAO) on sum of neutrino masses, as has been shown in figures \ref{f44},\ref{f84} and \ref{f104}.
	\item Figures \ref{f45} and \ref{f46} shows baryon asymmetry parameter $(\eta_{B})$ as a function of dark matter mass and sum  of neutrino masses respectively for weight $k_{Y}=4$ and obtained data is found to satisfy the observed value of the baryon asymmetry parameter while simultaneously satisfying the Lyman-$\alpha$ constraints on dark matter mass and Planck bound on sum of neutrino masses. 
	\item Figures \ref{f85} and \ref{f86} gives $\eta_{B}$ as a function of dark matter mass and sum of neutrino masses for weight $k_{Y}=8$. What has been observed here is that the Planck'18 limit on baryon asymmetry parameter is not satisfied, rather the data points are found to be greater than $10^{-7}$ and as such the result obtained does not define baryon asymmetry along with the study of dark matter. Also, from figure \ref{f86}, it has been observed that $\eta_{B}$ falls on the desired range only when the sum of neutrino masses is very small, around $10^{-12}$ to $10^{-10}$ for normal hierarchy and for inverted hierarchy it lies around the parameter space on and near $10^{-10}$.
	\item For weight $k_{Y}=10$, from figure \ref{f105} it can be inferred that for mass of the dark matter candidate lying within the allowed range, the observed value for $\eta_{B}$ is not satisfied for both normal and inverted hierarchy, however for baryon asymmetry parameter as a function of sum of neutrino masses, the results are found to be satisfactory as depicted in figure \ref{f106}.
\end{itemize}
Coming to the conclusive part of the work, so in the current scenario, modular symmetry has been incorporated into doublet LRSM, where the neutrino mass originates by the double seesaw mechanism. Taking into consideration the concept of multi-component dark matter, analysis for the same has been carried out for weights $k_{Y}=4,8$ and $10$. Also, considering that the decay of the right-handed neutrino $N_{1}$ gives rise to the desired CP asymmetry, a thorough analysis for leptogenesis has also been carried out for all the respective weights. Different results have been obtained for the varying weights and the results for both the hierarchies have been summarized using table,
\begin{table}[H]
	\begin{center}
		\begin{tabular}{|c|c|c|c|}
			\hline
			& $k_{Y}=4$(NH)[IH] & $k_{Y}=8$(NH)[IH] & $k_{Y}=10$(NH)[IH]\\
			\hline
			Dark Matter & $(\checkmark)[\checkmark]$ & $(\checkmark)[\checkmark]$ & $(\checkmark)[\checkmark]$ \\
			\hline
			Baryon asymmetry & $(\checkmark)[\checkmark]$ & $(\times)[\times]$ & $(\checkmark)[\checkmark]$ \\
			\hline
			Dark matter and baryon asymmetry & $(\checkmark)[\checkmark]$ & $(\times)[\times]$ & $(\times)[\times]$ \\
			\hline
		\end{tabular}
		\caption{\label{t6} The results summarized for dark matter and baryon asymmetry where $\checkmark$ denotes that the data points obtained give satisfactory results pertaining to experimental limits and $\times$ denote that the data obtained do not give the expected result. For dark matter, we consider the variation of $\Omega_{DM}h^{2}$ as a function of $\Sigma m_{\nu}$, for baryon asymmetry $\eta_{B}$ is considered as a function of $\Sigma m_{\nu}$ and for both together, variation of $\eta_{B}$ as a function of dark matter mass $m_{DM}$ is considered.}
	\end{center}
\end{table}
\begin{center}
	\section*{Appendix A : Properties of $A_4$ discrete symmetry group.}
\end{center}

$A_4$ is a non-abelian discrete symmetry group which represents even permuatations of four objects. It has four irreducible representations, three out of which are singlets $(1,1',1'')$ and one triplet $3$ ($3_A$ represents the anti-symmetric part and $3_S$ the symmetric part). Products of the singlets and triplets are given by,
\begin{center}
	\begin{equation*}
		1 \otimes 1 = 1
	\end{equation*}
\end{center}
\begin{center}
	\begin{equation*}
		1' \otimes 1' = 1''
	\end{equation*}
\end{center}
\begin{center}
	\begin{equation*}
		1' \otimes 1'' = 1
	\end{equation*}
\end{center}
\begin{center}
	\begin{equation*}
		1'' \otimes 1'' = 1'
	\end{equation*}
\end{center}
\begin{center}
	\begin{equation*}
		3 \otimes 3 = 1 \oplus 1' \oplus 1'' \oplus 3_A \oplus 3_S
	\end{equation*}
\end{center}
If we have two triplets under $A_4$ say, $(a_1,a_2,a_3)$ and $(b_1,b_2,b_3)$ , then their multiplication rules are given by,
\begin{center}
	\begin{equation*}
		1 \approx a_1b_1 + a_2b_3 + a_3b_2
	\end{equation*}
	\begin{equation*}
		1' \approx a_3b_3 + a_1b_2 + a_2b_1
	\end{equation*}
	\begin{equation*}
		1'' \approx a_2b_2 + a_3b_1 + a_1b_3
	\end{equation*}
	\begin{equation*}
		3_S \approx \begin{pmatrix}
			2a_1b_1-a_2b_3-a_3b_2 \\
			2a_3b_3-a_1b_2-a_2b_1 \\
			2a_2b_2-a_1b_3-a_3b_1
		\end{pmatrix}
	\end{equation*}
	\begin{equation*}
		3_A \approx \begin{pmatrix}
			a_2b_3-a_3b_2 \\
			a_1b_2-a_2b_1 \\
			a_3b_1-a_1b_3
		\end{pmatrix}
	\end{equation*}
\end{center}

\begin{center}
	\section*{Appendix B : Modular Symmetry.}
\end{center}

Modular symmetry has gained much importance in aspects of model building \cite{King:2020qaj}, \cite{Novichkov:2019sqv}. This is because it can minimize the extra particle called 'flavons' while analyzing a model with respect to a particular symmetry group. An element $q$ of the modular group acts on a complex variable $\tau$ which belongs to the upper-half of the complex plane given as \cite{Novichkov:2019sqv} \cite{Feruglio:2017spp} 
\begin{equation}
	\label{E:23}
	q\tau = \frac{a\tau+b}{c\tau+d}
\end{equation}
where $a,b,c,d$ are integers and $ad-bc=1$, Im$\tau$$>$0.\\
The modular group is isomorphic to the projective special linear group PSL(2,Z) = SL(2,Z)/$Z_2$ where, SL(2,Z) is the special linear group of integer $2\times2$ matrices having determinant unity and $Z_2=({I,-I})$ is the centre, $I$ being the identity element. The modular group can be represented in terms of two generators $S$ and $T$ which satisfies $S^2=(ST)^3=I$. $S$ and $T$ satisfies the following matrix representations:
\begin{equation}
	\label{E:24}
	S = \begin{pmatrix}
		0 & 1\\
		-1 & 0
	\end{pmatrix}
\end{equation}

\begin{equation}
	\label{e:25}
	T = \begin{pmatrix}
		1 & 1\\
		0 & 1
	\end{pmatrix}
\end{equation}

corresponding to the transformations,
\begin{equation}
	\label{e:26}
	S : \tau \rightarrow -\frac{1}{\tau} ; T : \tau \rightarrow \tau + 1
\end{equation}
Finite modular groups (N $\leq$ 5) are isomorphic to non-abelian discrete groups, for example, $\Gamma(3) \approx A_4$, $\Gamma(2) \approx S_3$, $\Gamma(4) \approx S_4$. While using modular symmetry, the Yukawa couplings can be expressed in terms of modular forms, and the number of modular forms present depends upon the level and weight of the modular form. For a modular form of level N and weight 2k, the table below shows the number of modular forms associated within and the non-abelian discrete symmetry group to which it is isomorphic \cite{Feruglio:2017spp}.
\begin{table}[H]
	\begin{center}
		\begin{tabular}{|c|c|c|}
			\hline
			N & No. of modular forms & $\Gamma(N)$ \\
			\hline
			2 & k + 1 & $S_3$ \\
			\hline
			3 & 2k + 1 & $A_4$ \\
			\hline
			4 & 4k + 1 & $S_4$ \\
			\hline
			5 & 10k + 1 & $A_5$ \\
			\hline 
			6 & 12k &  \\
			\hline
			7 & 28k - 2 & \\
			\hline
		\end{tabular}
		\caption{\label{table:3}No. of modular forms corresponding to modular weight 2k.}
	\end{center}
\end{table}
In our work, we will be using modular form of level 3, that is, $\Gamma(3)$ which is isomorphic to $A_4$ discrete symmetry group. The weight of the modular form is taken to be 2, and hence it will have three modular forms $(Y_1,Y_2,Y_3)$ which can be expressed as expansions of q given by,
\begin{equation}
	\label{e:27}
	Y_1 = 1 + 12 q + 36 q^2 + 12 q^3 + 84 q^4 + 72 q^5 + 36 q^6 + 96 q^7 + 
	180 q^8 + 12 q^9 + 216 q^{10}
\end{equation}
\begin{equation}
	\label{e:28}
	Y_2 = -6 q^{1/3} (1 + 7 q + 8 q^2 + 18 q^3 + 14 q^4 + 31 q^5 + 20 q^6 + 
	36 q^7 + 31 q^8 + 56 q^9)
\end{equation}
\begin{equation}
	\label{e:29}
	Y_3 = -18 q^{2/3} (1 + 2 q + 5 q^2 + 4 q^3 + 8 q^4 + 6 q^5 + 14 q^6 + 
	8 q^7 + 14 q^8 + 10 q^9)
\end{equation}
where, $q = \exp(2\pi i \tau)$.\\

\section*{Acknowledgement}

The author would like to acknowledge Professor Mrinal Kumar Das of Tezpur University for introducing the topic of Left-Right Symmetric Model and modular symmetry during the course of the author’s doctoral studies, as well as for his insightful discussions and valuable suggestions.

\bibliographystyle{unsrt}
\bibliography{citation2}

\begin{thebibliography}{10}

\bibitem{Zwicky:1933gu}
F.~Zwicky.
\newblock {Die Rotverschiebung von extragalaktischen Nebeln}.
\newblock {\em Helv. Phys. Acta}, 6:110--127, 1933.

\bibitem{Planck:2018vyg}
N.~Aghanim et~al.
\newblock {Planck 2018 results. VI. Cosmological parameters}.
\newblock {\em Astron. Astrophys.}, 641:A6, 2020.
\newblock [Erratum: Astron.Astrophys. 652, C4 (2021)].

\bibitem{Sakharov:1988vdp}
A.~D. Sakharov.
\newblock {Baryon asymmetry of the universe}.
\newblock pages 65--80, 1990.

\bibitem{Grimus:1993fx}
W.~Grimus.
\newblock {Introduction to left-right symmetric models}.
\newblock In {\em {4th Hellenic School on Elementary Particle Physics}}, pages
  619--632, 3 1993.

\bibitem{BhupalDev:2018xya}
P.~S. Bhupal~Dev, Rabindra~N. Mohapatra, Werner Rodejohann, and Xun-Jie Xu.
\newblock {Vacuum structure of the left-right symmetric model}.
\newblock {\em JHEP}, 02:154, 2019.

\bibitem{Lee:2017mfg}
Chang~Hun Lee.
\newblock {\em {Left-right symmetric model and its TeV-scale phenomenology}}.
\newblock PhD thesis, Maryland U., 2017.

\bibitem{Grimus:2013tva}
W.~Grimus and L.~Lavoura.
\newblock {Double seesaw mechanism and lepton mixing}.
\newblock {\em JHEP}, 03:004, 2014.

\bibitem{Maiezza:2016ybz}
Alessio Maiezza, Goran Senjanovi{\'c}, and Juan~Carlos Vasquez.
\newblock {Higgs sector of the minimal left-right symmetric theory}.
\newblock {\em Phys. Rev. D}, 95(9):095004, 2017.

\bibitem{Senjanovic:2016bya}
Goran Senjanovic.
\newblock {Is left{\textendash}right symmetry the key?}
\newblock {\em Mod. Phys. Lett. A}, 32(04):1730004, 2017.

\bibitem{Senjanovic:1978ev}
Goran Senjanovic.
\newblock {Spontaneous Breakdown of Parity in a Class of Gauge Theories}.
\newblock {\em Nucl. Phys. B}, 153:334--364, 1979.

\bibitem{Mohapatra:1977be}
Rabindra~N. Mohapatra and Deepinder~P. Sidhu.
\newblock {Gauge Theories of Weak Interactions with Left-Right Symmetry and the
  Structure of Neutral Currents}.
\newblock {\em Phys. Rev. D}, 16:2843, 1977.

\bibitem{Borah:2025fkd}
Debasish Borah, Satyabrata Mahapatra, Dibyendu Nanda, Sujit~Kumar Sahoo, and
  Narendra Sahu.
\newblock {Effective theory of light Dirac neutrino portal dark matter with
  observable {\ensuremath{\Delta}}Neff}.
\newblock {\em Phys. Rev. D}, 112(5):055010, 2025.

\bibitem{Das:2016akd}
Arindam Das, Natsumi Nagata, and Nobuchika Okada.
\newblock {Testing the 2-TeV Resonance with Trileptons}.
\newblock {\em JHEP}, 03:049, 2016.

\bibitem{Das:2017hmg}
Arindam Das, P.~S.~Bhupal Dev, and Rabindra~N. Mohapatra.
\newblock {Same Sign versus Opposite Sign Dileptons as a Probe of Low Scale
  Seesaw Mechanisms}.
\newblock {\em Phys. Rev. D}, 97(1):015018, 2018.

\bibitem{Patel:2023voj}
Utkarsh Patel, Pratik Adarsh, Sudhanwa Patra, and Purushottam Sahu.
\newblock {Leptogenesis in a Left-Right Symmetric Model with double seesaw}.
\newblock {\em JHEP}, 03:029, 2024.

\bibitem{Kakoti:2025eub}
Ankita Kakoti and Mrinal~Kumar Das.
\newblock {keV Sterile Neutrino as Dark Matter in Doublet Left-Right Symmetric
  Model with A4 Modular Symmetry}.
\newblock {\em PTEP}, 2025(12):123B12, 2025.

\bibitem{Patra:2023ltl}
Sudhanwa Patra, S.~T. Petcov, Prativa Pritimita, and Purushottam Sahu.
\newblock {Neutrinoless double beta decay in a left-right symmetric model with
  a double seesaw mechanism}.
\newblock {\em Phys. Rev. D}, 107(7):075037, 2023.

\bibitem{Dodelson:1993je}
Scott Dodelson and Lawrence~M. Widrow.
\newblock {Sterile-neutrinos as dark matter}.
\newblock {\em Phys. Rev. Lett.}, 72:17--20, 1994.

\bibitem{Boyarsky:2018tvu}
A.~Boyarsky, M.~Drewes, T.~Lasserre, S.~Mertens, and O.~Ruchayskiy.
\newblock {Sterile neutrino Dark Matter}.
\newblock {\em Prog. Part. Nucl. Phys.}, 104:1--45, 2019.

\bibitem{Drewes:2016upu}
M.~Drewes et~al.
\newblock {A White Paper on keV Sterile Neutrino Dark Matter}.
\newblock {\em JCAP}, 01:025, 2017.

\bibitem{Boubekeur:2023fqo}
Lotfi Boubekeur and Stefano Profumo.
\newblock {Tremaine-Gunn limit with mass-varying particles}.
\newblock {\em Phys. Rev. D}, 107(10):103535, 2023.

\bibitem{Davoudiasl:2020uig}
Hooman Davoudiasl, Peter~B. Denton, and David~A. McGady.
\newblock {Ultralight fermionic dark matter}.
\newblock {\em Phys. Rev. D}, 103(5):055014, 2021.

\bibitem{Merle:2015vzu}
Alexander Merle, Aurel Schneider, and Maximilian Totzauer.
\newblock {Dodelson-Widrow Production of Sterile Neutrino Dark Matter with
  Non-Trivial Initial Abundance}.
\newblock {\em JCAP}, 04:003, 2016.

\bibitem{Asaka:2006nq}
Takehiko Asaka, Mikko Laine, and Mikhail Shaposhnikov.
\newblock {Lightest sterile neutrino abundance within the nuMSM}.
\newblock {\em JHEP}, 01:091, 2007.
\newblock [Erratum: JHEP 02, 028 (2015)].

\bibitem{Gautam:2019pce}
Nayana Gautam and Mrinal~Kumar Das.
\newblock {Phenomenology of keV scale sterile neutrino dark matter with $S_{4}$
  flavor symmetry}.
\newblock {\em JHEP}, 01:098, 2020.

\bibitem{Davidson:2008bu}
Sacha Davidson, Enrico Nardi, and Yosef Nir.
\newblock {Leptogenesis}.
\newblock {\em Phys. Rept.}, 466:105--177, 2008.

\bibitem{Buchmuller:2005eh}
W.~Buchmuller, R.~D. Peccei, and T.~Yanagida.
\newblock {Leptogenesis as the origin of matter}.
\newblock {\em Ann. Rev. Nucl. Part. Sci.}, 55:311--355, 2005.

\bibitem{DiBari:2012fz}
Pasquale Di~Bari.
\newblock {An introduction to leptogenesis and neutrino properties}.
\newblock {\em Contemp. Phys.}, 53(4):315--338, 2012.

\bibitem{Nir:2007zq}
Yosef Nir.
\newblock {Introduction to leptogenesis}.
\newblock In {\em {6th Rencontres du Vietnam}: {Challenges in Particle
  Astrophysics}}, 2 2007.

\bibitem{Feruglio:2017spp}
Ferruccio Feruglio.
\newblock {\em {Are neutrino masses modular forms?}}, pages 227--266.
\newblock 2019.

\bibitem{King:2020qaj}
Simon J.~D. King and Stephen~F. King.
\newblock {Fermion mass hierarchies from modular symmetry}.
\newblock {\em JHEP}, 09:043, 2020.

\bibitem{Sahu:2020tqe}
Purushottam Sahu, Sudhanwa Patra, and Prativa Pritimita.
\newblock {$A_4$ realization of left-right symmetric linear seesaw}.
\newblock 2 2020.

\bibitem{Ding:2022aoe}
Gui-Jun Ding, F.~R. Joaquim, and Jun-Nan Lu.
\newblock {Texture-zero patterns of lepton mass matrices from modular
  symmetry}.
\newblock {\em JHEP}, 03:141, 2023.

\bibitem{Lu:2019vgm}
Jun-Nan Lu, Xiang-Gan Liu, and Gui-Jun Ding.
\newblock {Modular symmetry origin of texture zeros and quark lepton
  unification}.
\newblock {\em Phys. Rev. D}, 101(11):115020, 2020.

\bibitem{Zhang:2019ngf}
Di~Zhang.
\newblock {A modular $A_4$ symmetry realization of two-zero textures of the
  Majorana neutrino mass matrix}.
\newblock {\em Nucl. Phys. B}, 952:114935, 2020.

\bibitem{Blanchet:2008pw}
Steve Blanchet and Pasquale Di~Bari.
\newblock {New aspects of leptogenesis bounds}.
\newblock {\em Nucl. Phys. B}, 807:155--187, 2009.

\bibitem{Novichkov:2019sqv}
P.~P. Novichkov, J.~T. Penedo, S.~T. Petcov, and A.~V. Titov.
\newblock {Generalised CP Symmetry in Modular-Invariant Models of Flavour}.
\newblock {\em JHEP}, 07:165, 2019.

\end{thebibliography}

\end{document}